\documentclass[aps,pre,preprint,groupedaddress]{revtex4-1}
\usepackage{epsf}
\usepackage{colordvi}
\usepackage{color}
\usepackage{graphicx}
\usepackage{fancybox}
\usepackage{times}                 
\usepackage{amssymb}      
\usepackage{epsf}
\usepackage{amsmath}
\usepackage{scrextend}
\usepackage[normalem]{ulem}
\usepackage{tikz}

\usepackage{framed}

\newcommand{\cs}[1]{\color{black}#1\normalcolor}
\begin{document}


\title{Swimming Speeds of Filaments in Viscous Fluids with Resistance}


\author{Nguyenho Ho}
\email[]{nho@wpi.edu}
\thanks{NSF DMS 1413110 }
\affiliation{Department of Mathematical Sciences, Worcester Polytechnic Institute, 100 Institute Road Worcester, MA 01609 USA}

\author{Karin Leiderman}
\email[]{kleiderman@ucmerced.edu}
\thanks{NSF DMS 1413078}
\affiliation{Applied Mathematics Unit, University of California Merced, 5200 N Lake Road, Merced, CA 95343 USA}

\author{Sarah D. Olson}
\email[]{sdolson@wpi.edu}
\homepage[]{http://users.wpi.edu/$\sim$sdolson/}
\thanks{NSF DMS 1413110}
\affiliation{Department of Mathematical Sciences, Worcester Polytechnic Institute, 100 Institute Road Worcester, MA 01609 USA}

\date{\today}

\begin{abstract}
\cs{Many microorganisms swim in a highly heterogeneous environment with obstacles such as fibers or polymers. To better understand how this environment affects microorganism swimming, we study propulsion of a cylinder or filament in a fluid with a sparse, stationary network of obstructions modeled by the Brinkman equation. The mathematical analysis of swimming speeds is investigated by studying an infinite-length  cylinder propagating lateral or spiral displacement waves. For fixed bending kinematics, we find that swimming speeds are enhanced due to the added resistance from the fibers. In addition, we examine the work and the torque exerted on the cylinder in relation to the resistance. The solutions for the torque, swimming speed, and work of an infinite-length cylinder in a Stokesian fluid are recovered as the resistance is reduced to zero. Finally, we compare the asymptotic solutions with the numerical results obtained from the Method of Regularized Brinkmanlets. The swimming speed of a finite-length filament decreases as its length decreases and planar bending induces an angular velocity that increases linearly with added resistance. The comparisons between the asymptotic analysis and computation give insight on the effect of the length of the filament, the permeability, and the thickness of the cylinder in terms of the overall performance of planar and helical swimmers. }
\end{abstract}

\pacs{}
\keywords{swimming microorganisms, Brinkman, resistance, porous medium }

\maketitle

\newpage

\section{Introduction}
\label{sec:intro}

%

The self-propulsion of microorganisms that utilize flagellar propulsion has been  the topic of a vast number of analytical, experimental, and computational studies for many years (reviewed in \cite{Lauga09}). Many species of spermatozoa and bacteria are able to swim by propagating lateral or spiral waves  along their cylindrical flagella \cite{Brennen77,Woolley01,Smith09Surf}. Similarly, larger organisms such as \textit{C. elegans} (nematodes) are also able to make forward progression through soil via undulatory locomotion \cite{Gagnon13}. The native environment in which these organisms live varies greatly. For example, spermatozoa  encounter different fluid environments in the female reproductive tract that include swimming through or around mucus, cells, hormones, and other large proteins \cite{Fauci06,Rutllant05,Suarez06}. Similarly, bacteria are able to swim in the mucus layer that coats the stomach and move in biofilms with extracellular polymeric substances \cite{Brennen77,Celli09,Flemming10}.

One may wonder, how does the swimming speed or mode of swimming change in these different environments?
Early experiments showed that Leptospira, a slender helical bacterium, is able to swim faster in methylcellulose (MC), a gel with chains of long polymers \cite{Berg79}. Another study showed that swimming speeds of seven different types of bacteria were enhanced in higher viscosity solutions of MC and PVP (polyvinylpyrollidone) \cite{Schneider74}; beyond a certain viscosity or polymer concentration, this enhancement was no longer observed. Experiments of sperm in MC and PA (polyacrylamide) gels showed that swimming speeds, beat frequency, and amplitude of undulation vary as the viscosity and concentration of the gels are varied \cite{Smith09b,Suarez92}. \textit{C. elegans} have also been observed to swim faster in polymer networks \cite{Gagnon13}.

Since the length scale of these swimmers is small, they live in a viscosity dominated environment where inertia can be neglected. Many studies have focused on analyzing idealized swimmers in viscous fluids at zero Reynolds number. Seminal work by GI Taylor examined swimming speeds of an infinite sheet in two-dimensions (2D) and an infinite cylinder with circular cross section of small radius in three-dimensions (3D), propagating lateral displacement waves \cite{Taylor51,Taylor52}. In these studies, it was shown that the second order swimming speed scales quadratically with amplitude and linearly with frequency for small amplitude bending. This analysis has been extended for several different cases including swimming speeds for cylinders with non-circular cross sections \cite{Kosa10}, as well as improvements to the perturbation series \cite{Sauzade12}. 

Since the fluid that these swimmers are moving through contains different amounts of proteins or other structures, more complex fluid models have been proposed and analyzed. For the case of a swimming sheet, studies have looked at the asymptotic swimming speeds in a gel represented as a two-phase fluid (elastic polymer network and viscous fluid) where enhancement in propulsion was observed for stiff and compressible networks \cite{Fu10}. In contrast, a two-fluid model (with intermixed fluids) exhibited a decreased swimming speed relative to the case of a fluid with a single viscosity in both asymptotics and numerical simulations \cite{Du12}. In another model, Magariyama \textit{et al.} \cite{Magariyama02} looked at a fluid governed by two viscosities using a modified resistive force theory and found that there is an enhancement in propulsion efficiency when the viscosity of the polymer solution increases and the other fluid viscosity is held constant. Swimming in a shear thinning fluid has also been studied; locomotion of finite-length swimmers is enhanced (2D numerical simulations) \cite{Montenegro13} and infinite undulating sheets exhibit a decrease in swimming speed relative to the Stokes case \cite{Dasgupta13}. 

Since the proteins or polymer chains in gels may cause a fluid to exhibit a nonlinear strain response (frequency dependent), viscoelastic fluid models have also been considered. Through asymptotic analysis, it has been shown that the swimming speed of infinite sheets and cylinders in viscoelastic fluids decreases relative to the speed in a purely viscous fluid \cite{Fu07,Fu09,Lauga07}. Simulations of finite-length swimmers in a viscoelastic fluid at zero Reynolds number governed by the Oldroyd B equation revealed that enhancement in swimming speeds can be observed when asymmetrical beatforms and swimmer elasticity work together \cite{Teran10,Thomases14}. Specifically, increases in swimming speeds were observed in a viscoelastic fluid when the beat frequency of the swimmer is on the same time scale as the polymer relaxation time \cite{Teran10}; when the polymer relaxation time is fast, other models may be more appropriate to understand swimming speeds. 

Another approach is to think of the fluid with an embedded polymer network as a porous medium. Darcy's law has been used to describe the fluid flow in porous media, where average velocity is proportional to the gradient in pressure. 
This law is not able to capture contributions of the viscous stress tensor and it is only valid on the macroscopic scale where the domain is large and boundary effects can be neglected \cite{Brinkman47,Koplik83}. To overcome these disadvantages, the incompressible Brinkman  flow equation has an  additional diffusion term \cite{Brinkman47}, 
\begin{eqnarray}
\nabla p=\mu\Delta {\bf u}-\dfrac{\mu}{\gamma}{\bf u},\hspace{0.5cm}\nabla\cdot{\bf u}=0, \label{BrEq}
\end{eqnarray}
where $p$ is the pressure, ${\bf u}$ is the velocity of the fluid, $\gamma$ is the permeability of the porous medium, and the effective viscosity is $\mu$. This equation represents the effective flow through a network of stationary obstacles with small volume fraction \cite{Auriault09,Brinkman47,Howells74,Spielman}. The resistance due to the obstacles is characterized by $\mu/\gamma$. Note that the incompressible Stokes equations are recovered in the limit as $\gamma\to \infty$, \cs{and when $\gamma\rightarrow0$, Eq.~(\ref{BrEq}) will behave like Darcy's law.  Another characteristic of a Brinkman fluid is the Brinkman screening length, $\sqrt{\gamma}$, which marks the approximate length over which a disturbance to the velocity would decay. For comparison, in 3D, the flow due to a point force in Stokes flow decays as $1/r$ whereas the flow due to a point force in a Brinkman flow decays like ${\gamma}/r^3$ \cite{Durlofsky87, Leshansky09}.} 

In the case of a two-phase fluid composed of a polymer network and solvent, if the polymer is stationary, we obtain the Brinkman equation. In this limiting case of a two-phase fluid, an infinite-length sheet exhibits an enhancement in swimming speed \cite{Fu10}. Previously, Leshansky \cite{Leshansky09} derived the asymptotic swimming speed for an infinite sheet propagating waves of lateral bending in a fluid governed by the Brinkman equation. They observed that swimming speeds scaled similarly to those of Stokes, scaling quadratically with amplitude. In addition to the Stokesian swimming speed, there is an extra factor that depends on the permeability and is monotonically increasing for decreasing permeability (increasing the resistance in the fluid). 

In this paper, we focus on calculating the asymptotic swimming speed for a waving cylindrical tail that exhibits lateral displacement waves in a Brinkman fluid. 
A second order asymptotic swimming speed is derived for planar bending 
and we find that swimming speeds are enhanced, similar to the 2D case for an infinite sheet. Swimming speeds are also calculated for cylindrical tails with spiral displacement waves, showing that fluid resistance enhances swimming speed. These results shed insight on how added fluid resistance changes propulsion of cylindrical tails when the kinematics are prescribed. In addition, as the resistance approaches zero, we recover the swimming speed, work, and torque for an infinite-length cylinder in a fluid governed by the Stokes equation. \cs{ Through our analysis, we also find the range of enhancement in swimming speeds for the infinite cylinder in a Brinkman fluid and the relation to permeability, cylinder thickness, and wavenumber.} \cs{To validate our asymptotic results, we apply the method of Regularized Brinkmanlets to study finite-length filaments. Through validation, we find that the theoretical swimming speed of filaments with planar bending waves matches up well with the simulation data and that the asymptotics overestimate swimming speeds for shorter length cylindrical swimmers. In the helical bending wave case, we calculate the external torque exerted on the filament by the surrounding fluid. We observe that the numerical and the asymptotic findings may not consistently agree with one another; the asymptotics overestimate the torque of finite-length helical swimmers. }

\section{Swimming speeds for a cylinder with planar bending}
\label{sec:model}
\begin{figure}[!htb]
\centering
\includegraphics*[width=.4\textwidth]{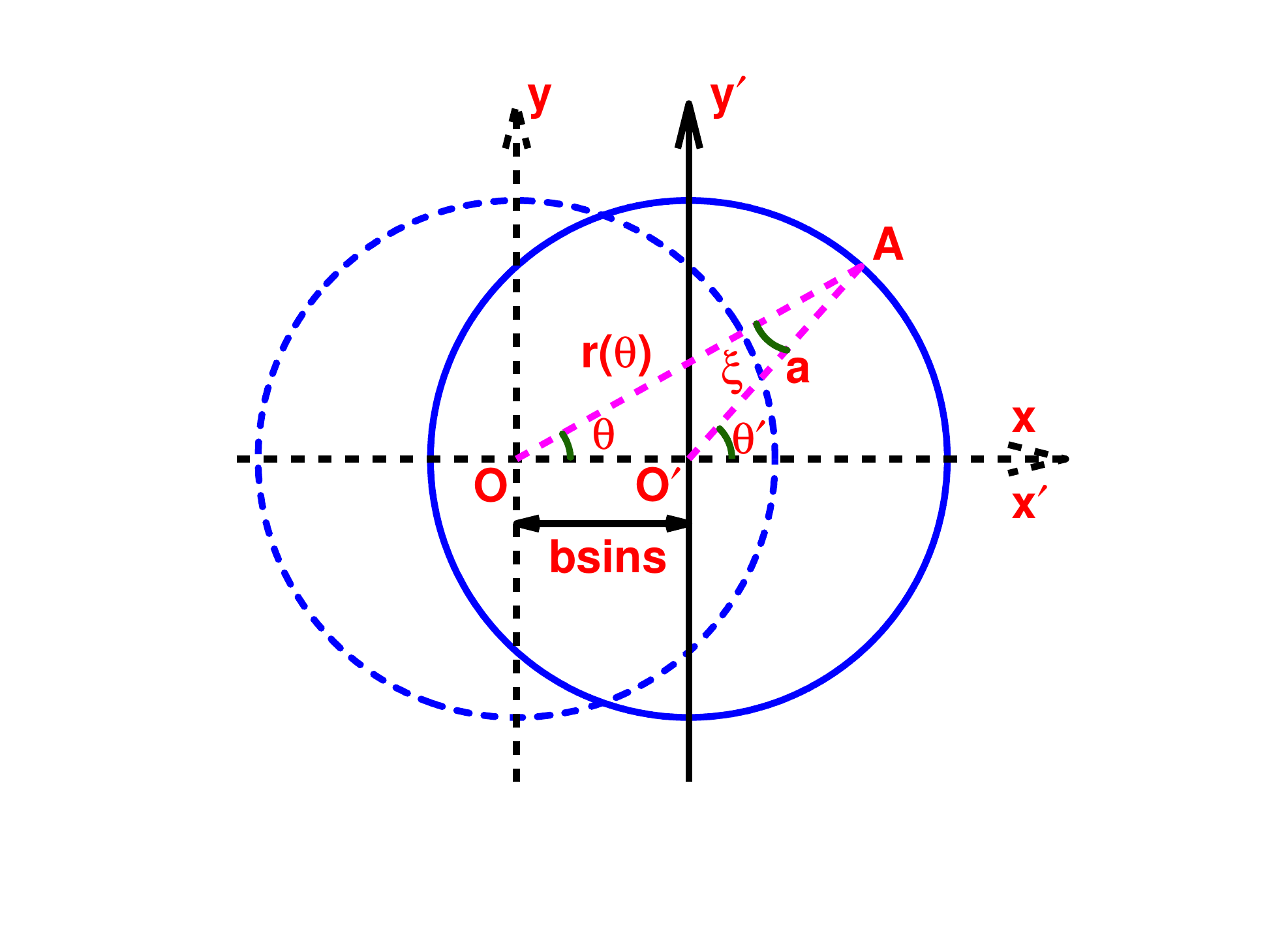}
\caption{Current configuration (deformed state) of the cylinder propagating bending waves is shown with the solid circle. $O'$ is the origin for the current configuration and $O$ is the center of the original (non-deformed) state that is shown with the dashed circle. }
\label{CylGeomPic}
\end{figure}

\subsection{Cylinder with lateral displacement waves }
Similar to previous work \cite{Fu09,Taylor52}, we consider a cylinder of constant cross section, bending with small amplitude in the $x$ direction, immersed in a fluid. The cylinder is bending in the $x-y$ direction with
\begin{eqnarray*}
x=b\sin (k(z+Ut)), \hspace{2mm} y=0,
\end{eqnarray*}
where $b$ is the amplitude, $U$ is the velocity of the propagating wave, and $k$ is the wavenumber, defined as $k=2\pi/\lambda$ where $\lambda$ is the wavelength. With this, the velocity components of the cylinder have the form $u_x=bkU\cos( k(z+Ut))$ and $u_y=0$.
To simplify, we let $s=k(z+Ut)$ and 
convert the above equations into cylindrical coordinates to obtain the 
boundary conditions on the surface of the cylinder,
\begin{eqnarray}
u_r=bkU\cos\theta\cos s, \hspace{2mm} u_{\theta}=-bkU\sin\theta\cos s, \hspace{2mm} u_z=0 .\label{velBC}
\end{eqnarray}
From this point, we will regard the velocity components in cylindrical coordinates as $u_r=u, u_{\theta}=v$ and $u_z=w$. 

The time-dependent position of the cylinder at any given point on the surface is given as
\begin{eqnarray}
r^2=a^2+b^2\sin^2s+2ab\sin s\cos\theta',\label{rEq}
\end{eqnarray}
as shown in Fig.~\ref{CylGeomPic} for $\theta'=\theta+\xi$. As detailed in Appendix \ref{Rderive}, we can show that
\begin{eqnarray}
r&=&a+b\sin s\cos\theta\label{shapeBC}
\end{eqnarray}
in the first order of $b/a$. 

\subsection{Fluid Model}
The 3D Brinkman equation in cylindrical coordinates is: 
\begin{eqnarray}
\dfrac{1}{\mu}\dfrac{\partial p}{\partial r}&=&\dfrac{\partial}{\partial r}\left[\dfrac{1}{r}\dfrac{\partial}{\partial r}(ru)\right]+\dfrac{1}{r^2}\dfrac{\partial^2u}{\partial\theta^2}-\dfrac{2}{r^2}\dfrac{\partial v}{\partial\theta}+\dfrac{\partial^2u}{\partial z^2}-\dfrac{1}{\gamma}u\label{radialeq}\\ 
\dfrac{1}{\mu r}\dfrac{\partial p}{\partial \theta}&=&\dfrac{\partial}{\partial r}\left[\dfrac{1}{r}\dfrac{\partial}{\partial r}(rv)\right]+\dfrac{1}{r^2}\dfrac{\partial^2v}{\partial\theta^2}+\dfrac{2}{r^2}\dfrac{\partial u}{\partial\theta}+\dfrac{\partial^2v}{\partial z^2}-\dfrac{1}{\gamma}v\label{tangentialeq}\\ 
\dfrac{1}{\mu}\dfrac{\partial p}{\partial z}&=&\dfrac{1}{r}\dfrac{\partial}{\partial r}\left(r\dfrac{\partial w}{\partial r}\right)+\dfrac{1}{r^2}\dfrac{\partial^2 w}{\partial\theta^2}+\dfrac{\partial^2 w}{\partial z^2}-\dfrac{1}{\gamma}w\label{adialeq}
\end{eqnarray}
where $u$, $v$ and $w$ are the velocity components in the direction of $r$, $\theta$, and $z$, respectively. The continuity equation for the incompressible flow is given by
\begin{eqnarray}
\dfrac{\partial u}{\partial r}+\dfrac{u}{r}+\dfrac{1}{r}\dfrac{\partial v}{\partial\theta}+\dfrac{\partial w}{\partial z}=0.\label{coneq}
\end{eqnarray}
Taking the divergence of Eq.~\eqref{radialeq} and using Eq.~\eqref{coneq} to simplify, we find that the pressure satisfies $\nabla^2 p=0$. 
Let $\zeta=kr$ and recall $s=k({z}+Ut)$. The general solution for the pressure is thus
 \begin{eqnarray}
p=\mu knA_{mn}K_{m}(n\zeta)\cos(m\theta)\cos(ns),\label{press}
\end{eqnarray}
where $K_{m}$ is the $m^{th}$ order modified Bessel function of the second kind and $A_{mn}$ is a constant which is evaluated using the boundary conditions \cite{Happel83}. Based on  the pressure in Eq.~\eqref{press}, we assume the  velocity components can be described as
 \begin{eqnarray}
 u=u_{mn}\cos m\theta\cos ns, \hspace{2mm v=v_{mn}\sin m\theta \cos ns}, \hspace{2mm}\mbox{and}\hspace{2mm} w=w_{mn}\cos m\theta \sin ns. \label{vel3}
 \end{eqnarray}
Note that $u_{mn}, v_{mn}$ and $w_{mn}$ are functions with respect to $\zeta$ only. Substituting $u$, $v$, $w$, and $p$ from Eqs.~\eqref{press}--\eqref{vel3} into Eqs.~\eqref{radialeq}--\eqref{tangentialeq} and using the relations $s=k(z+Ut)$ and $\zeta=kr$, we obtain the following system of equations:
 \begin{eqnarray}
\left[\dfrac{\partial^2}{\partial \zeta^2}+\dfrac{1}{\zeta}\dfrac{\partial}{\partial \zeta}-\dfrac{(m+1)^2}{\zeta^2}-\left(n^2+\dfrac{\alpha^2}{k^2}\right)\right](u_{mn}+v_{mn})=-n^2A_{mn}K_{m+1}(n\zeta)\label{eq1},\\ 
\left[\dfrac{\partial^2}{\partial \zeta^2}+\dfrac{1}{\zeta}\dfrac{\partial}{\partial \zeta}-\dfrac{(m-1)^2}{\zeta^2}-\left(n^2+\dfrac{\alpha^2}{k^2}\right)\right](u_{mn}-v_{mn})=-n^2A_{mn}K_{m-1}(n\zeta)\label{eq2},
\end{eqnarray}
with $\alpha^2=1/\gamma$ (where $\gamma$ is the Darcy permeability). \cs{The parameter $\alpha$ is known as the hydrodynamic resistance of the porous medium and has units of inverse length. In addition, $\alpha$ is proportional to the ratio of the diameter of the fiber over the spacing within the network. This ratio is usually characterized as the mesh spacing \cite{Durlofsky87}.}

The homogeneous solutions for Eqs.~\eqref{eq1}--\eqref{eq2} include the modified Bessel function of the first kind, which will diverge as $\zeta\rightarrow \infty$. Thus, we eliminate this solution to maintain finite values for the velocities. The particular solutions are 
  \begin{eqnarray}
(u_{mn}+v_{mn})_p=\dfrac{A_{mn}}{\beta^2}n^2K_{m+1}(n\zeta) \hspace{5mm}\text{and}\hspace{5mm} (u_{mn}-v_{mn})_p=\dfrac{A_{mn}}{\beta^2}n^2K_{m-1}(n\zeta),
\end{eqnarray}
\cs{where $\beta=\alpha/k$ is the scaled resistance. It is a nondimensional constant that characterizes the relationship between the resistance or average mesh size and the wavelength of the swimmer}. After simplifying, the general solutions to Eqs.~\eqref{eq1}--\eqref{eq2} are
 \begin{eqnarray}
u_{mn}+v_{mn}&=&B_{mn}K_{m+1}(\chi \zeta)+\dfrac{A_{mn}}{\beta^2}n^2K_{m+1}(n\zeta),\label{uvplus}\\
u_{mn}-v_{mn}&=&C_{mn}K_{m-1}(\chi \zeta)+\dfrac{A_{mn}}{\beta^2}n^2K_{m-1}(n\zeta),\label{uvminus}
\end{eqnarray}
for $\chi=\sqrt{n^2+\beta^2}$. The constants $B_{mn}$ and $C_{mn}$ are determined by the boundary conditions of the cylindrical tail. The radial and tangential velocity components are found to satisfy the following equations:
 \begin{eqnarray}
2u_{mn}&=&B_{mn}K_{m+1}(\chi \zeta)+C_{mn}K_{m-1}(\chi \zeta)+\dfrac{A_{mn}n^2}{\beta^2}\left[K_{m+1}(n\zeta)+K_{m-1}(n\zeta)\right],\label{ueq}\\
2v_{mn}&=&B_{mn}K_{m+1}(\chi \zeta)-C_{mn}K_{m-1}(\chi \zeta)+\dfrac{2A_{mn}}{\beta^2\zeta}mnK_m(n\zeta).\label{veq}
\end{eqnarray}
The axial component of the velocity is determined using the continuity condition given in Eq.~\eqref{coneq} and is given by
 \begin{eqnarray}
nw_{mn}&=&-\left[\dfrac{\partial u_{mn}}{\partial \zeta}+\dfrac{1}{\zeta}\left(u_{mn}+mv_{mn}\right)\right]\nonumber\\
&=&\dfrac{A_{mn}n^3}{\beta^2}K_{m}(n\zeta)+\dfrac{1}{2}(B_{mn}+C_{mn})\chi K_m(\chi \zeta).\label{weqn}   \end{eqnarray}
 \cs{Since our goal is to determine the swimming speed of the cylinder, we will have to determine the first and second order solutions, using the condition that the disturbance caused by the cylinder body should vanish at infinity \cite{Taylor52}.}

  \subsection{First order solution}
 As detailed in Appendix \ref{asymptoticexpand}, the velocity components are expanded about $\zeta=\zeta_1=ka$. To the first order, when $m=1$ and $n=1$ , the boundary conditions are $u_1=bkU$, $v_1=-bkU$, and $w_1=0$. Plugging into Eqs.~\eqref{uvplus}-\eqref{uvminus} and \eqref{weqn}, we obtain:
   \begin{eqnarray}
u_1+v_1&=&BK_2(\chi \zeta_1)+\dfrac{A}{\beta^2}K_2(\zeta_1)=0,\label{u1plusv1}\\
u_1-v_1&=&CK_0(\chi \zeta_1)+\dfrac{A}{\beta^2}K_0(\zeta_1)=2bkU\label{u1minusv1},\\
w_1&=&\dfrac{A}{\beta^2}K_1(\zeta_1)+\dfrac{1}{2}(B+C)\chi K_1(\chi \zeta_1)=0\label{w1eq},
   \end{eqnarray}
for $\chi=\sqrt{1+\beta^2}$.   From Eqs.~\eqref{u1plusv1}-\eqref{w1eq}, the constants are
\begin{eqnarray}
\dfrac{A}{\beta^2}&=&-\dfrac{1}{\phi(\zeta_1)}\dfrac{2bkU}{K_0(\chi \zeta_1)},\label{Abeta}\\
B&=&\dfrac{1}{\phi(\zeta_1)}\dfrac{2bkU}{K_0(\chi \zeta_1)}\dfrac{K_2(\zeta_1)}{K_2(\chi \zeta_1)},\label{BBC}\\
C&=&\dfrac{2bkU}{K_0(\chi \zeta_1)}\left[1+\dfrac{1}{\phi(\zeta_1)}\dfrac{K_0(\zeta_1)}{K_0(\chi \zeta_1)}\right],\label{CBC}
   \end{eqnarray}
 where
      \begin{eqnarray}
\phi(\zeta_1)=\dfrac{2}{\chi}\dfrac{K_1(\zeta_1)}{K_1(\chi \zeta_1)}-\dfrac{K_0(\zeta_1)}{K_0(\chi \zeta_1)}-\dfrac{K_2(\zeta_1)}{K_2(\chi \zeta_1)}.\label{phizeta}
   \end{eqnarray}    
     \cs{To determine the velocity of the cylinder, we have that Eqs.~(\ref{u1u2})-(\ref{w1w2}) in Appendix \ref{asymptoticexpand} will vanish at infinity \cite{Taylor52}. Thus, there is no contribution to the swimming speed of the cylinder in the first order expansion.}
      \subsection{Second order solution}
The second order expansions and boundary conditions are detailed in Appendix \ref{asymptoticexpand}. Using the same argument for the velocity of the filament at infinity, we arrive at 
  \begin{eqnarray*}
U_{\infty}=\dfrac{1}{4}bk(w_1')_{\zeta=\zeta_1=ka},\label{Vspeed}
   \end{eqnarray*}
   where $w_1'$ is the first derivative of the axial velocity component given in Eq.~\eqref{weqn} with respect to $\zeta$ (for $m=1, n=1$). Using the first order solution, and evaluating $w_1'$ at the boundary,  $\zeta=\zeta_1=ka$, we have 
 \begin{eqnarray}
w_1'
&=&\dfrac{2bkU}{\phi(\zeta_1)}\left[\dfrac{K_0(\zeta_1)}{K_0(\chi \zeta_1)}-\chi\dfrac{K_1(\zeta_1)}{K_1(\chi \zeta_1)}\right].\label{w1d}
   \end{eqnarray}
The swimming speed $U_{\infty}$ up to second order expansion is thus
            \begin{eqnarray}
U_{\infty}&=&\dfrac{1}{2}b^2k^2U\dfrac{1}{\phi(\zeta_1)}\left[\dfrac{K_0(\zeta_1)}{K_0(\chi \zeta_1)}-\chi\dfrac{K_1(\zeta_1)}{K_1(\chi \zeta_1)}\right]. \label{Vspeed2}
   \end{eqnarray}
The asymptotic velocity for an infinite-length cylinder that is propagating planar bending waves in a Brinkman fluid is given above in Eq.~\eqref{Vspeed2} and depends on the scaled resistance $\alpha/k$ through $\chi$.

\begin{figure}[!htb]
\centering
\includegraphics*[width=.47\textwidth]{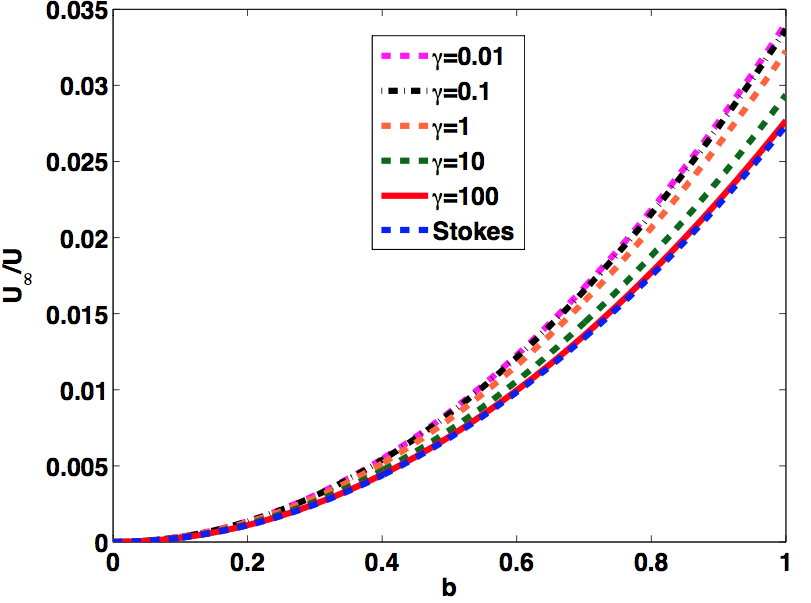}
\includegraphics*[width=.49\textwidth]{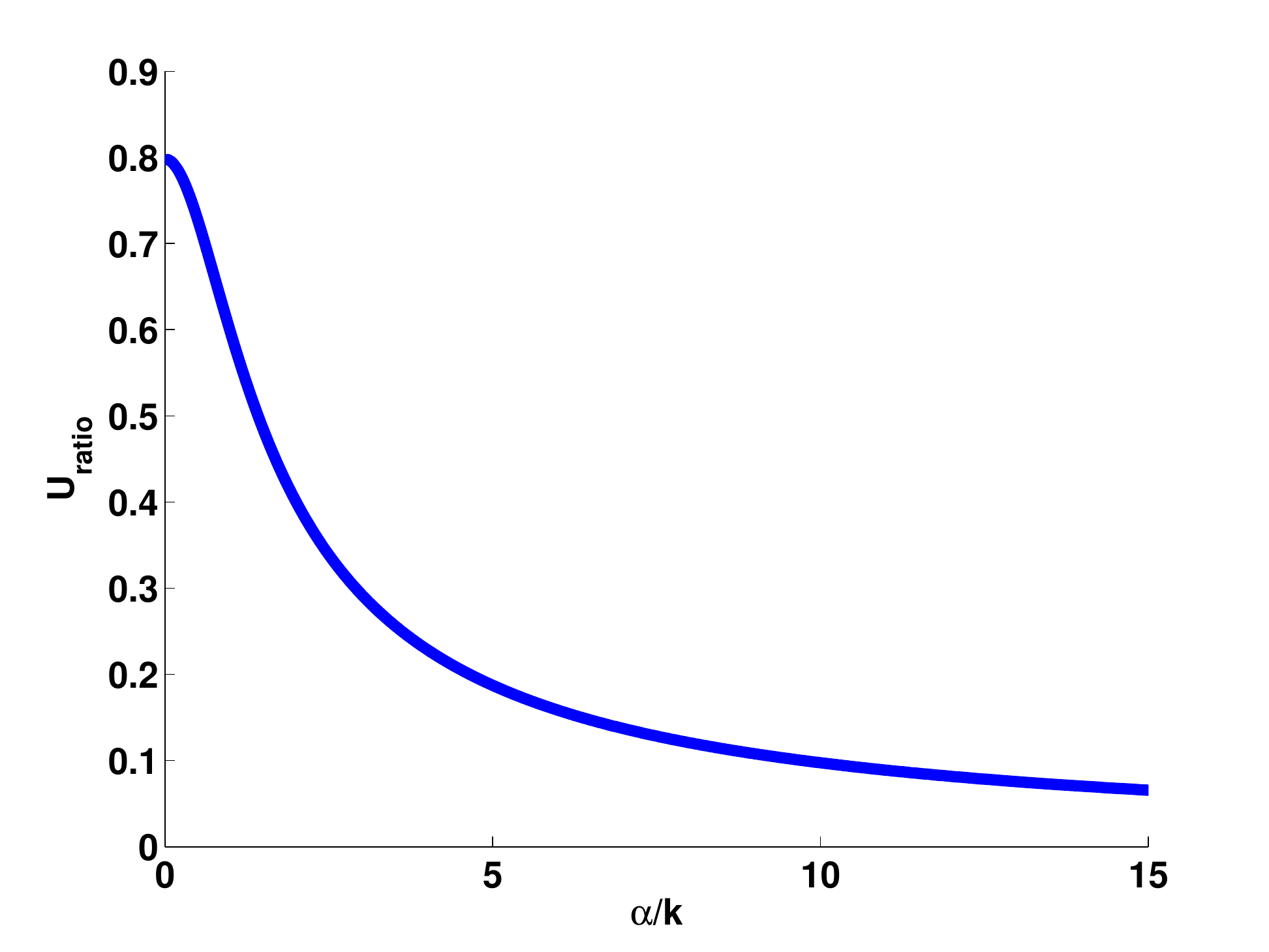}\\
{\bf (a)} \hspace{16em} {\bf (b)}
\caption{(a) The nondimensional swimming speed of a cylinder with planar undulations, calculated from Eq.~\eqref{velsmallr}, is shown for several permeability values $\gamma$ for fixed wavelength $\lambda=24$  and $a=0.05$. The Stokes case is also plotted for comparison. \cs{(b) The ratio between the swimming speed in Eq.~\eqref{velsmallr} with the speed of a swimming sheet derived from \cite{Leshansky09} for different scaled resistance $\alpha/k$ and with $k=2\pi/24$.} }\label{swimplanar}
\end{figure}

 \cs{In the limiting case when $\zeta_1=ka\ll 1$, the limit forms of the Bessel functions are \cite{Milton72}}:
\cs{\begin{eqnarray*}
K_1(\zeta_1)=\dfrac{1}{\zeta_1}+\mathcal{O}(\zeta_1)&,& \hspace{5mm} K_1(\chi \zeta_1)=\dfrac{1}{\chi \zeta_1}+\mathcal{O}(\chi \zeta_1),\\
K_2(\zeta_1)=\dfrac{2}{\zeta_1^2}+\mathcal{O}(\zeta_1^2)&,&\hspace{5mm} K_2(\chi \zeta_1)=\dfrac{2}{\chi^2\zeta_1^2}+\mathcal{O}(\chi^2 \zeta_1^2),\\
K_0(\zeta_1)=-\log\zeta_1-\gamma_e+\log2+\mathcal{O}(\zeta_1^2)&,& \hspace{0.4cm} K_0(\chi \zeta_1)=-\log \zeta_1+\log2-\gamma_e-\log\chi+\mathcal{O}(\chi^2\zeta_1^2),\label{BessEG}
   \end{eqnarray*}
where $\gamma_e$ is the Euler-Mascheroni constant. Thus, for $\zeta_1=ka\ll 1$ we can rewrite $\phi(\zeta_1)$ as
    \begin{eqnarray*}
\phi(\zeta_1)=2-\chi^2-\dfrac{K_0(\zeta_1)}{K_0(\zeta_1)-\log\chi}.
   \end{eqnarray*}}
   To second order, the nondimensional swimming speed, $U_{\infty}/U$, in the case of a cylinder propagating lateral bending waves is given as
            \begin{equation}
\dfrac{U_{\infty}}{U}=\dfrac{1}{2}b^2k^2\left[\dfrac{(1-\chi^2)K_0(\zeta_1)+\chi^2\log\chi}{(1-\chi^2)K_0(\zeta_1)-(2-\chi^2)\log\chi}\right],\label{velsmallr}
   \end{equation}
for $ka\ll 1$. We note that this swimming speed scales quadratically with the amplitude of bending $b$ and depends on the resistance $\alpha$ through the parameter $\chi$. The swimming speeds are shown in Fig.~\ref{swimplanar} for several permeability values $\gamma$. For comparison, we also plot the swimming speed of the same infinite-length cylinder propagating planar bending in a fluid governed by the incompressible Stokes equation, as derived by Taylor \cite{Taylor52}. We observe in Fig.~\ref{swimplanar} that as $\alpha\to0$ (or $\gamma\to\infty$), we approach the Stokes swimming speed. In the next section, we will study this case further. 
%
%
 \subsection{Comparison of Swimming Speeds} 
 The Brinkman equation  is characterized by the Darcy permeability $\gamma$. In the case of $\gamma\to\infty$ (or resistance $\alpha\to0$), we recover the Stokes equation. To understand what happens to the swimming speed of the infinite-length cylinder (with $ka\ll 1$) as $\alpha\to0$, we will work with Eq.~\eqref{velsmallr} to obtain the following expression,
  \begin{eqnarray}
\dfrac{U_{\infty}}{U}&=&\dfrac{1}{2}b^2k^2\left[\dfrac{K_0(\zeta_1)-\dfrac{1}{2}\log\left(1+\dfrac{\alpha^2}{k^2}\right)^{\dfrac{k^2}{\alpha^2}}-\dfrac{1}{2}\log\left(1+\dfrac{\alpha^2}{k^2}\right)}{K_0(\zeta_1)+\dfrac{1}{2}\log\left(1+\dfrac{\alpha^2}{k^2}\right)^{\dfrac{k^2}{\alpha^2}}-\dfrac{1}{2}\log\left(1+\dfrac{\alpha^2}{k^2}\right)}\right]\label{velasymexpand}.
   \end{eqnarray}
We note the following limits as $\alpha\to0$:
 \begin{eqnarray}
\lim_{\alpha\rightarrow0}\log\left(1+\dfrac{\alpha^2}{k^2}\right)^{\dfrac{k^2}{\alpha^2}}=1, \hspace{15mm} \lim_{\alpha\rightarrow0}\log\left(1+\dfrac{\alpha^2}{k^2}\right)=0.\label{limit_ind}
   \end{eqnarray}
   Thus,  the second order asymptotic velocity of a cylinder with $\zeta_1= ka\ll 1$ in a Brinkman fluid becomes
    \begin{eqnarray*}
\dfrac{U_{\infty}}{U}=\dfrac{1}{2}b^2k^2\left[\dfrac{K_0(\zeta_1)-\dfrac{1}{2}}{K_0(\zeta_1)+\dfrac{1}{2}}\right].
   \end{eqnarray*}
 This is precisely the asymptotic velocity of the same cylinder immersed in a fluid governed by the Stokes equations as derived by Taylor \cite{Taylor52}.
 
\cs{Next, we study the swimming speed of the infinite-length 3D cylinder in comparison to the 2D sheet, where both are propagating planar bending waves. The propulsion of an undulating planar sheet was studied by Leshansky \cite{Leshansky09} and the swimming speed $U_{Les}$ was found to be:
        \begin{eqnarray}
\dfrac{U_{Les}}{U}&=&\dfrac{1}{2}b^2k^2\sqrt{1+\alpha^2/k^2},\label{LesPlanarSheet}
   \end{eqnarray}
for $\alpha^2=1/{\gamma}$. The ratio of Eq.~(\ref{velasymexpand}) and (\ref{LesPlanarSheet}) is
        \begin{eqnarray}
U_{ratio}=\dfrac{U_{\infty}}{U_{Les}}&=&\left[\dfrac{K_0(\zeta_1)-\dfrac{1}{2}\left(\dfrac{k^2}{\alpha^2}+1\right)\log\left(1+\dfrac{\alpha^2}{k^2}\right)}{K_0(\zeta_1)+\dfrac{1}{2}\left(\dfrac{k^2}{\alpha^2}-1\right)\log\left(1+\dfrac{\alpha^2}{k^2}\right)}\right]\cdot\dfrac{1}{\sqrt{1+\dfrac{\alpha^2}{k^2}}}.\label{URatioPlot}
   \end{eqnarray}
We plot $U_{ratio}$ versus the scaled resistance $\alpha/k$ in Fig.~\ref{swimplanar}(b). We observe that the ratio decreases as $\alpha$ increases. This implies that the 3D infinite-length cylinder swims slower than the 2D sheet in a fluid with the same Darcy permeability. When $\alpha/k\rightarrow0$, we see that the ratio approaches 
    \begin{eqnarray*}
U_{ratio}=\dfrac{K_0(\zeta_1)-1/2}{K_0(\zeta_1)+1/2},\label{Stratio}
   \end{eqnarray*}
for a fixed $\zeta_1$. This is the ratio of the swimming speeds of the infinite-length 3D cylinder and 2D sheet  in a fluid governed by the Stokes equation.}

 \section{Energy to maintain planar bending}

The force on the surface is calculated as ${\bf F}=\boldsymbol{\sigma}\cdot{\bf n}$ where $\boldsymbol{\sigma}$ is the stress tensor and ${\bf n}$ is the normal vector. The velocity components of ${\bf u}$ at the boundary $r=a$ are given in Eq.~(\ref{velBC}). The stress tensor components are given by $\sigma_{rr}=-p+2\mu\dfrac{\partial u}{\partial r}$ and $\sigma_{r\theta}=\mu\left(\dfrac{\partial v}{\partial r}-\dfrac{v}{r}+\dfrac{1}{r}\dfrac{\partial u}{\partial \theta}\right)$. Evaluating the stress using the first order velocity solution at the boundary gives  
\begin{eqnarray*}
\left(\sigma_{rr}\right)_{\zeta=\zeta_1}&=&\mu k\left[-AK_1(\zeta_1)\right]\cos\theta\cos s,\\
\left(\sigma_{r\theta}\right)_{\zeta=\zeta_1}&=&\mu k\left(\dfrac{\partial v_1}{\partial \zeta}\right)_{\zeta=\zeta_1}\sin\theta\cos s,
\end{eqnarray*}
where $r=\zeta/k$. \cs{Since we consider a fluid with a low volume fraction of stationary and randomly oriented fibers, the total stress applied to the filament is assumed to be entirely due to the fluid and not influenced by the fibers. This assumption is valid since the distance between the fibers is large compared to the radius of the filament. There is further discussion of this in Section \ref{rangeassumption}}. 

\begin{figure}[!htb]
\centering
\includegraphics*[width=.5\textwidth]{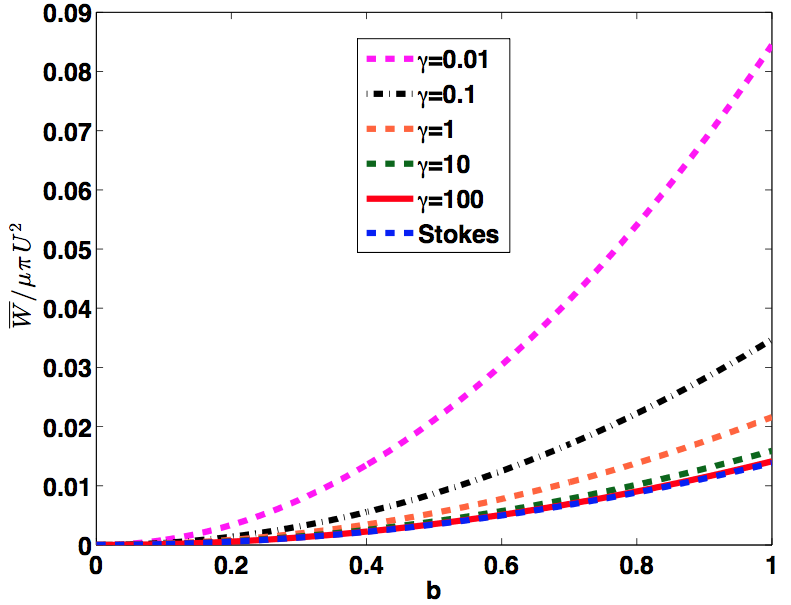}
\caption{The nondimensional rate of mean work done to maintain planar bending along the infinite-length cylinder, calculated from Eq.~\eqref{Wrad} for several permeabilities $\gamma$ where $\lambda=24$ and $a=0.05$. The Stokes case is also plotted for comparison.}\label{fig:work}
\end{figure}

The rate of work done to maintain planar bending is calculated as follows:
\begin{equation}
dW=-{\bf F}\cdot{\bf u}=\mu bk^2U\left[AK_1(\zeta_1)\cos^2\theta+\left(\dfrac{\partial v_1}{\partial \zeta}\right)_{\zeta=\zeta_1}\sin^2\theta\right]\cos^2s.\label{ratework1}
\end{equation} 
Using Eq.~\eqref{veq}, the derivative of $v_1$ is:
\begin{eqnarray}
\dfrac{\partial v_1}{\partial \zeta}&=&\dfrac{1}{2}(-B+C)\chi K_1(\chi \zeta)-\dfrac{1}{\zeta}\left[\dfrac{A}{\beta^2}K_2(\zeta)+BK_2(\chi \zeta)\right],\label{v1d_eva}
\end{eqnarray}
where $\dfrac{A}{\beta^2}, B, C$ are from Eqs.~(\ref{Abeta})-(\ref{CBC}).
The mean value of the rate of work to maintain the filament motion is denoted by $\overline{dW}$ and is given as
\begin{eqnarray*}
\overline{dW}&=&\dfrac{1}{4}\mu bk^2U\left\{\dfrac{2bkU}{\zeta_1\phi(\zeta_1)}\left[-\dfrac{\beta^2}{K_0(\chi \zeta_1)}-\dfrac{\chi^2}{2K_0(\chi \zeta_1)}+\dfrac{1}{2}\dfrac{\phi(\zeta_1)}{K_0(\chi \zeta_1)}+\dfrac{1}{2}\dfrac{K_0(\zeta_1)}{K_0^2(\chi \zeta_1)}\right]\right\}.
\end{eqnarray*}
For a cylinder immersed in a Brinkman fluid, the mean value of the total rate of work per unit length ($\lambda$) along the surface of the cylinder ($r=a$) is then calculated as
\begin{eqnarray}
\overline{W}&=&\dfrac{\mu\pi b^2k^2U^2}{K_0(\zeta_1)+\dfrac{1}{2}\left(\dfrac{k^2}{\alpha^2}-1\right)\log\left(1+\dfrac{\alpha^2}{k^2}\right)},\label{Wrad}
\end{eqnarray}
where $\phi(\zeta_1)=2-\chi^2-\dfrac{K_0(\zeta_1)}{K_0(\chi \zeta_1)}$ when $\zeta_1$ is small, and $\chi=\sqrt{1+\beta^2}$. When the permeability approaches infinity, the Brinkman fluid behaves like Stokes flow. Thus, when $\gamma\to\infty$ (or $\alpha\to0$) and using Eq.~(\ref{limit_ind}), we have
\begin{eqnarray*}
\overline{W}&=& \dfrac{\mu\pi b^2k^2U^2}{K_0(\zeta_1)+1/2}.
\end{eqnarray*}
This is exactly the same energy contribution to maintain the flagellum in motion  in Stokes flow \cite{Taylor52}. The nondimensional rate of work is shown in Fig.~\ref{fig:work} for several different permeabilities $\gamma$ and we observe that as $\gamma$ gets large, it approaches the work done in a Stokesian fluid. \cs{ In this analysis, as the permeability decreases, we observe that there are small changes in the swimming speed (shown in Fig.~\ref{swimplanar}(a)), but the work done increases greatly (shown in Fig.~\ref{fig:work}). The mathematical analysis for this observation is detailed in Appendix \ref{AsymptoticFuncs}. The physical meaning of this phenomenon can be explained as follows. For a small permeability, there is a large added resistance present in the fluid, preventing the swimmer from propelling itself forward. Therefore, it requires more work to move with the same prescribed kinematics. We note that the rate of work of the swimming sheet has been previously calculated and is also an increasing function of resistance $\alpha$  \cite{Leshansky09}. }

  \section{Cylinder with spiral bending}
\begin{figure}[!htb]
\centering
\includegraphics*[width=.4\textwidth]{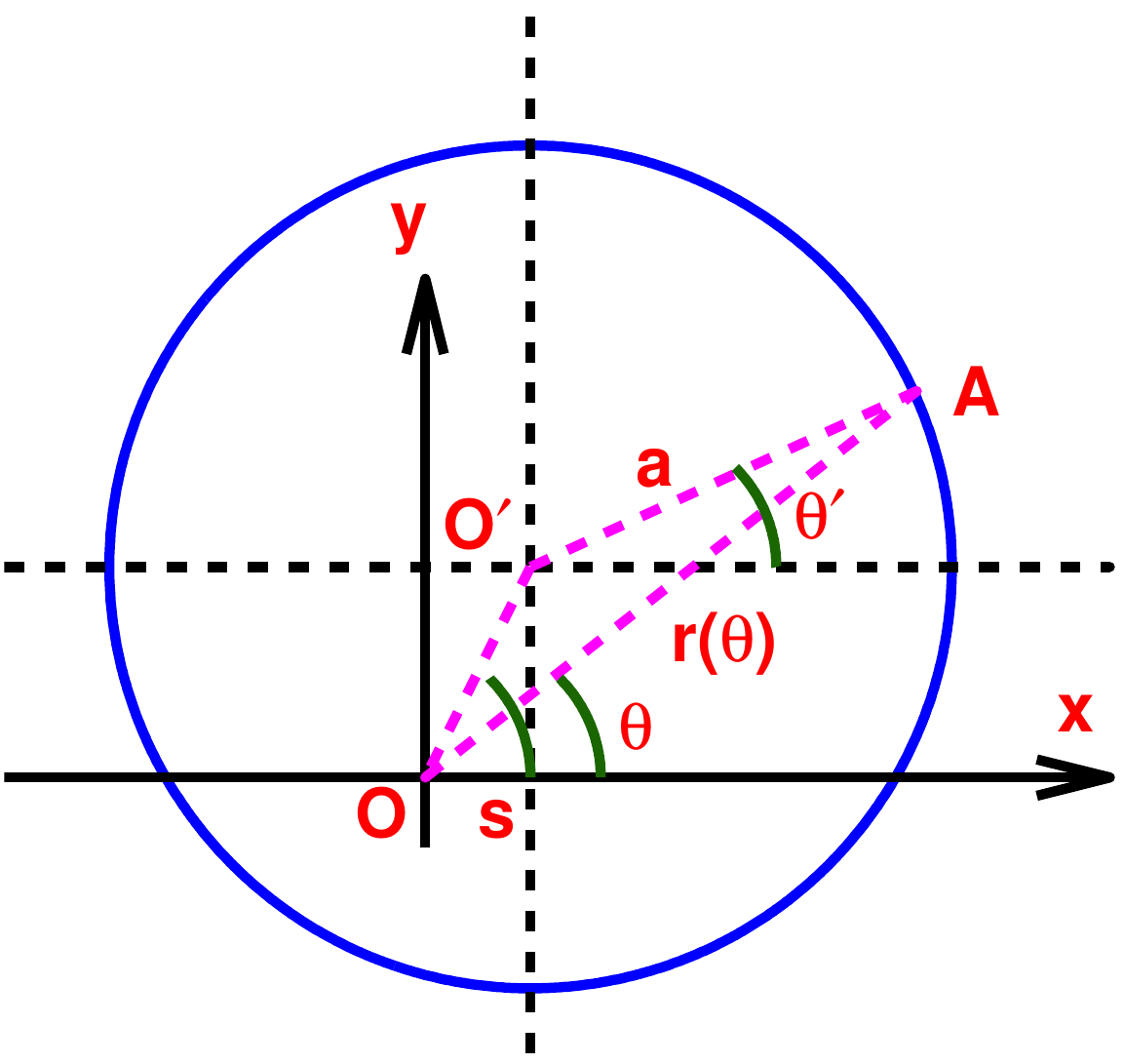}
\caption{Model geometry for a cylinder propagating spiral bending waves. The circle is the cross section of the deformed cylinder (current configuration) which is centered at $O'$. The undeformed cylinder is centered at $O$.}
\label{SpiralGeomPic}
\end{figure}

Next, we consider an infinite-length cylinder propagating spiral waves, motivated by experiments where sperm flagella are able to exhibit helical bending \cite{Woolley01}. Thus, it is compelling to consider the rotational movements of a cylinder propagating spiral bending waves (helical bending waves with constant radius). One can verify from Fig.~\ref{SpiralGeomPic} that
similar to the planar case, to the first order of $b/a$,
\begin{eqnarray}
r=a+b\cos(\theta-s),\label{rsp}
\end{eqnarray}
where details can be found in Appendix \ref{RderiveSpiral} and $s=k(z+Ut)$.
 Eq.~\eqref{rsp} corresponds to a cylinder that will achieve the form of a right-handed helix about its axis with angular velocity $kU$ in the direction of increasing $\theta$. The formulation for the cylinder is 
\begin{eqnarray*}
x=b\cos s, \hspace{2mm} y=b\sin s, \hspace{2mm} z=bs,
\end{eqnarray*}
and the velocity components become
\begin{eqnarray*}
u_x=-bkU\sin s, \hspace{2mm} u_y=bkU\cos s, \hspace{2mm} u_z=0.
\end{eqnarray*}
Converting the above equations to cylindrical coordinates, we have
\begin{eqnarray}
u=bkU\sin(\theta-s), \hspace{2mm} v=bkU\cos(\theta-s), \hspace{2mm} w=0 .\label{spiralBC}
\end{eqnarray}

The motion of the helix includes the contributions of two orthogonal planar motions that are perpendicular to the $z$-axis, namely the $xz$-plane and $yz$-plane. The analysis for each plane  proceeds in a similar fashion to that of the planar case, satisfying the boundary conditions in Eq.~(\ref{spiralBC}). As previous analysis has shown, the second-order solution can only be determined through first-order expansions (see \cite{Kosa10,Taylor52}). The second-order velocity components at the boundary are 
\begin{eqnarray}
u_{22}=-\dfrac{1}{2}bk(u'_1)_{\zeta=\zeta_1}, \hspace{5mm} v_{22}=-\dfrac{1}{2}bk(v'_1)_{\zeta=\zeta_1}, \hspace{5mm} w_{22}=-\dfrac{1}{2}bk(w'_1)_{\zeta=\zeta_1}.\label{spbc}
\end{eqnarray}
Let $V_s$ be the propulsion velocity of the helix in the opposite direction of the propagating spiral bending waves. With this, similar to \cite{Taylor52}, we have
\begin{eqnarray*}
V_s=\dfrac{1}{2}bk\left(w'_1\right)_{\zeta=\zeta_1} 
\end{eqnarray*}
where $w'_1$ is the same as in Eq.~(\ref{w1d}). By a simple calculation, we observe  
\begin{eqnarray}
\dfrac{V_s}{U}=b^2k^2\cdot\dfrac{K_0(\zeta_1)-\chi^2K_0(\chi \zeta_1)}{(2-\chi^2)K_0(\chi \zeta_1)-K_0(\zeta_1)}.\label{SpiralVel}
\end{eqnarray}
Similar to the results obtained in the planar case, when $\alpha\to0$, we recover the speed $V_s$ in the incompressible Stokes equations,
\begin{eqnarray*}
\dfrac{V_s}{U}=b^2k^2\cdot\dfrac{K_0(\zeta_1)-1/2}{K_0(\zeta_1)+1/2}.
\end{eqnarray*}
Thus, the swimming speed of a spiral bending wave is double that of a planar bending wave with the same kinematics. \cs{We note that modified resistive force theory calculations have also been used to determine expressions for the swimming speed of a spiral bending wave in a Brinkman fluid \cite{Leshansky09}.}

In addition to determining the asymptotic swimming speed from spiral bending, we can find the expression for the torque exerted on the cylinder by the surrounding fluid. Since the fluid in this case flows in a circular motion, the radial and axial velocity components are zero and only tangential velocity plays a role in this calculation. That is,
\begin{eqnarray*}
u=0, \hspace{10mm} v=\dfrac{\Omega}{r}=\dfrac{\Omega k}{\zeta}, \hspace{10mm} w=0,
\end{eqnarray*}
where $\Omega$ is the angular velocity of the helix. With this,  we simplify the expression for mean torque per unit length applied on the filament by the fluid, $T_{\infty}=2\pi r^2\sigma_{r\theta}$, 
to
%
\begin{eqnarray}
T_{\infty}=-4\pi\mu \Omega.\label{MeanTorqueBC}
\end{eqnarray}
To solve for $\Omega$, we use the boundary condition for $v_{22}$ in Eq.~\eqref{spbc} to obtain
\begin{eqnarray}
\dfrac{\Omega k}{\zeta_1}=-\dfrac{1}{2}bk(v'_1)_{\zeta=\zeta_1}.
\label{angBC}
\end{eqnarray}
Substituting Eq.~(\ref{angBC}) into Eq.~(\ref{MeanTorqueBC}) and using Eq.~(\ref{v1d_eva}) for $v'_1$ and  Eqs.~(\ref{Abeta})-(\ref{CBC}) to simplify, we have
\begin{eqnarray}
T_{\infty}=\dfrac{4\pi\mu b^2kU}{K_0(\zeta_1)+\dfrac{1}{2}\left(\dfrac{k^2}{\alpha^2}-1\right)\log\left(1+\dfrac{\alpha^2}{k^2}\right)}\label{torqueanalysis}.
\end{eqnarray}
In the limit as $\alpha\to0$, the torque exerted on the cylinder reduces to
$$T_{\infty}=\dfrac{4\pi\mu b^2kU}{K_0(\zeta_1)+1/2},$$
which is the same torque calculated for the Stokes regime by Drummond \cite{Drummond66}. Note that this derivation differs from the work of Taylor \cite{Taylor52} (where $w_1'$ was used instead of $v_1'$).
\cs{\section{Range of Parameters That Lead To Swimming Speed Enhancement}
To identify the range of parameter values that lead to enhancement in swimming speeds of the infinite-length cylinder with planar waves, we rearrange Eq.~(\ref{velasymexpand}) as follows:
 \begin{small}
    \begin{equation}
\dfrac{U_{\infty}}{U}=\dfrac{1}{2}b^2k^2\dfrac{K_0(\zeta_1)-\dfrac{1}{2}}{K_0(\zeta_1)+\dfrac{1}{2}}\left\{1+\dfrac{K_0(\zeta_1)-\dfrac{1}{2}\log\left(1+\dfrac{\alpha^2}{k^2}\right)-K_0(\zeta_1)\dfrac{k^2}{\alpha^2}\log\left(1+\dfrac{\alpha^2}{k^2}\right)}{\left[K_0(\zeta_1)-\dfrac{1}{2}\right]\left[K_0(\zeta_1)+\dfrac{1}{2}\left(\dfrac{k^2}{\alpha^2}-1\right)\log\left(1+\dfrac{\alpha^2}{k^2}\right)\right]}\right\}.\label{extendvel}
   \end{equation}
      \end{small}
%
Note that Eq.~(\ref{extendvel}) illustrates the velocity behavior in the spiral bending wave case when the constant $1/2$ is removed. The swimming speed is increasing when the following inequalities hold:
    \begin{eqnarray}
    K_0(\zeta_1)>\dfrac{\dfrac{1}{2}\dfrac{\alpha^2}{k^2}\log\left(1+\dfrac{\alpha^2}{k^2}\right)}{\dfrac{\alpha^2}{k^2}-\log\left(1+\dfrac{\alpha^2}{k^2}\right)},\label{K0equation}\\
  \zeta_1=ka<\dfrac{2}{e^{\gamma_e}}\exp\left\{-\dfrac{\dfrac{1}{2}\dfrac{\alpha^2}{k^2}\log\left(1+\dfrac{\alpha^2}{k^2}\right)}{\dfrac{\alpha^2}{k^2}-\log\left(1+\dfrac{\alpha^2}{k^2}\right)}\right\}=h\left(\dfrac{\alpha}{k}\right).\label{zetaineq}
   \end{eqnarray}
Therefore, for any fixed permeability, the swimming speed of the cylinder is enhanced if the thickness and wavelength of the cylinder satisfies the inequality in (\ref{zetaineq}). 
\begin{figure}[!htb]
\centering
\includegraphics*[width=.5\textwidth]{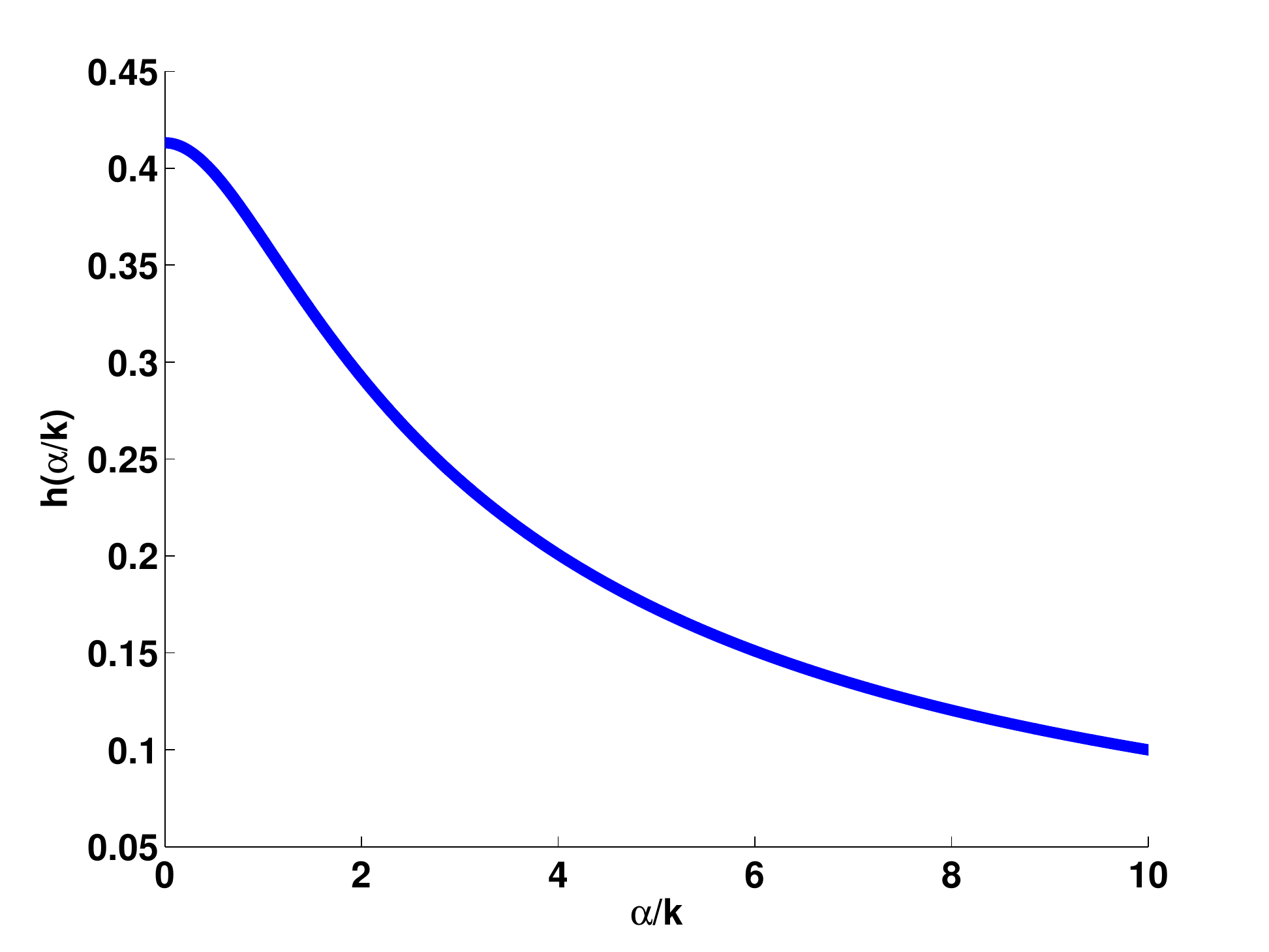}
\caption{\cs{The plot of the function on the right hand side of Eq.~(\ref{zetaineq}). }}
\label{VelWorkFunction}
\end{figure}
In Fig.~\ref{VelWorkFunction}, we plot the right hand side  of Eq.~(\ref{zetaineq}), $h(\alpha/k)$, to show that it is, in fact, decreasing in a manner that is dependent on the scaled resistance. This means that if the permeability is reduced, then $ka$ must also be reduced to observe swimming enhancement in a Brinkman fluid.
Hence, the cylinder radius and/or wavelength must decrease in order to observe an increase in swimming speed. This finding makes sense since the mesh size decreases as the permeability decreases, thus there is less room for the swimmer to move. We note that in addition to an enhancement in swimming speed, an increase in torque and rate of work will also be observed when Eq.~\eqref{zetaineq} is satisfied.}

\cs{\section{Range of Permeability and Swimming Enhancement}
\label{rangeassumption}
Our assumption is that the effective fluid environment can be modeled as a viscous fluid moving through a porous, static network of fibers via the Brinkman equation. For small volume fractions of fibers, this assumption is thought to be a valid one \cite{Auriault09}. Further, for randomly oriented fibers, Spielman and Goren \cite{Spielman} have derived a relationship between the volume fraction $\phi$, the permeability $\gamma$, and the radius of the fiber $a_{f}$, as
   \begin{eqnarray}
\dfrac{a_{f}^2}{\gamma}=4\phi\left[\dfrac{1}{3}\dfrac{a_{f}^2}{\gamma}+\dfrac{5}{6}\dfrac{a_{f}}{\sqrt{\gamma}}\dfrac{K_1(a_{f}/\sqrt{\gamma})}{K_0(a_{f}/\sqrt{\gamma})}\right].\label{fibrouseq}
   \end{eqnarray}
Since the Brinkman model assumes that the fiber network is static, we must have that the distance between the fibers (or the interfiber spacing) is large enough for the swimmer to move through with little or no interaction with the fibers. To estimate the ratio of interfiber spacing and the fiber radius, we use the following equation \cite{Leshansky09}:
   \begin{eqnarray}
\dfrac{D}{a_{f}}\approx2\left(\dfrac{1}{2}\sqrt{\dfrac{3\pi}{\phi}}-1\right),\label{ratiointerfiber}
   \end{eqnarray}
where $D$ is interfiber spacing. 
In the case where this ratio is large, there are little or no interactions between a stationary network of fibers and the swimmers. Thus, it is assumed that the fibers do not impart any additional stress onto the filament. 

In Table~\ref{table:CalculatedPermeability}, we report a few parameter ranges in which we see enhancement of swimming speed. In particular, we report ranges of the cylinder radius $a$, with a fixed wavelength of $\lambda=25~\mu$m. To find these ranges, we use fiber volume fractions and radii from the literature \cite{saltzman1994antibody}, together with our own computed values of permeability from Eq.~(\ref{fibrouseq}) and average separation from Eq.~(\ref{ratiointerfiber}).

\begin{table}[tbh!]
\centering
\begin{tabular}{c|c|c|c|c|c|c}
\hline\hline
Media~&~$\phi$ ~&~ $a_{f}$ (nm) ~&~ $D$ (nm) ~&~  $\gamma$ ($\mu$m$^2$) ~&~Eq.~(\ref{zetaineq}), $\lambda=25\mu$m~\\
\hline
Collagen gels, \cite{saltzman1994antibody}		&~ 0.00074 ~	&75	&8314& 8.6 &$~a<1.337 ~(\mu m)$\\
Cervical mucus, \cite{saltzman1994antibody}	~&0.015	&15	&346& 0.0085 &$~a<0.102~(\mu m)$\\
\hline 
\end{tabular}
\caption{The calculated  permeability $\gamma$ using the given volume fractions $\phi$ and fiber radii $a_{f}$. The range of cylinder thickness $a$ where an enhancement in swimming speed is observed is reported.}
\label{table:CalculatedPermeability}
\end{table}
The radii of the principal piece of human, bull, and ram sperm are $0.5~\mu$m, $0.29~\mu$m and $0.15~\mu$m, respectively \cite{hafez2012atlas,bahr1964study,bloodgood2013ciliary}. We note that that flagellar radius decreases along the length of the flagellum from the principal piece (closer to cell body) to the endpiece. Thus, swimmers will experience enhancement when placed in a collagen gel. However, there will be no enhancement for the three swimmers when they are put in cervical mucus at a volume fraction of $\phi=0.015$. Further, it is well known that the composition of the cervical and vaginal fluid varies greatly through the menstrual or oestrous cycle \cite{Aguilar05,Rutllant05}, and this experimental value of $\phi=0.015$ is taken at one time point in the cycle \cite{saltzman1994antibody}. For instance, around the time of ovulation, the interfiber spacing $D$ may reach up to $25$ $\mu$m \cite{Rutllant05}. Using this interfiber spacing and a given fiber radius $a_{f}=15$ nm, we can further estimate the volume fraction $\phi=0.00033$ from Eq.~\eqref{ratiointerfiber} and the permeability $\gamma=0.868$ ($\mu$m$^2$) from Eq.~\eqref{fibrouseq}. Then, the cylinder radii for which enhancement is seen is $a<0.765$ $\mu$m when the wavelength $\lambda$ is 25 $\mu$m. At this volume fraction, all three spermatozoa species will experience an enhancement in swimming speed in cervical fluid. }

\section{Numerical Studies}
\label{sec:numstud}
\subsection{ Background}\label{sec:regBrinkman}
\cs{In this section, we verify our asymptotic solutions and explore aspects of finite-length swimmers using the Method of Regularized Brinkmanlets (MRB) \cite{Cortez10}. This method is an extensiton of the Method of Regularized Stokeslets developed by Cortez \cite{Cortez01,Cortez05} for use with the Stokes equations. The general idea is to compute regularized fundamental solutions by replacing singular point forces with a smooth approximation. With this, the resulting equations can be solved exactly to obtain non-singular fundamental solutions. The smooth approximations to a delta distribution, often called `blob' functions, are characterized by a small parameter $\varepsilon$ that controls their width. The singular solutions can be recovered by letting $\varepsilon\rightarrow0$.
%

The Brinkmanlet is the fundamental solution to the singularly forced Brinkman equation 
\begin{eqnarray}
  \nabla p=\mu\Delta {\bf u}-\dfrac{\mu}{\gamma}{\bf u}+{\bf f}_0\delta({\bf x}-{\bf x}_0),\label{BrinkmanForce}
   \end{eqnarray}
   where ${\bf x}$ is any point in the fluid, ${\bf x}_0$ is the point where the force is applied, and $\delta(\cdot)$ is the delta distribution.
The pressure and velocity are in the form \cite{Cortez10}:
  \begin{eqnarray}
p&=&{\bf f}_0\cdot\nabla G({\bf x}-{\bf x}_0),\label{pBrinkman}\\
\mu{\bf u}&=&{\bf f}_0\cdot\nabla\nabla B({\bf x}-{\bf x}_0)-{\bf f}_0\Delta B({\bf x}-{\bf x}_0),\label{velBrinkman}
   \end{eqnarray}
   where $G(r)$ is the Green's function and $B(r)$ is related to $G(r)$ by the non-homogeneous Helmholtz differential equation $(\Delta -\alpha^2)B(r)=G(r)$ for $r=\|{\bf x}-{\bf x}_0\|$ and $\alpha^2=1/{\gamma}$. The solutions of $G(r)$ and $B(r)$ are well known \cite{Pozrikidis1989a,Cortez10}:
     \begin{eqnarray}
G(r)=-\dfrac{1}{4\pi r}, \hspace{10mm} B(r)=\dfrac{1-e^{-\alpha r}}{4\pi \alpha^2 r},
   \end{eqnarray}
and, thus, the Brinkmanlet velocity in (\ref{velBrinkman}) becomes
        \begin{eqnarray}
\mu{\bf u}({\bf x})={\bf f}_0H_1(r)+({\bf f}_0\cdot({\bf x}-{\bf x}_0))({\bf x}-{\bf x}_0)H_2(r),\label{RegVelBrink}
   \end{eqnarray}
  where $H_1(r)$ and $H_2(r)$ are functions of $G(r)$, $B(r)$, and their derivatives.
To regularize the fundamental solution, the expression for $B(r)$ is rewritten as $$B_{\varepsilon}(r)=\dfrac{1-e^{-\alpha R}}{4\pi \alpha^2 R},$$
where $R^2=r^2+\varepsilon^2$ so that the singularity is removed. 
From \cite{Cortez10}, the regularized Brinkmanlet velocity is 
         \begin{eqnarray}
\mu{\bf u}({\bf x})={\bf f}_0H_1^{\varepsilon}(r)+({\bf f}_0\cdot({\bf x}-{\bf x}_0))({\bf x}-{\bf x}_0)H_2^{\varepsilon}(r),\label{RegVelBrink}
   \end{eqnarray}
where
       \begin{eqnarray}
H_2^{\varepsilon}(r)&=&\dfrac{3}{4\pi\alpha^2R^5}-\dfrac{e^{-\alpha R}}{4\pi R^3}\left(\dfrac{3}{\alpha^3R^3}+\dfrac{3}{\alpha R}+1\right),\\
H_1^{\varepsilon}(r)&=&-\dfrac{1}{4\pi\alpha^2R^3}+\dfrac{e^{-\alpha R}}{4\pi R}\left(\dfrac{1}{\alpha^2R^2}+\dfrac{1}{\alpha R}+1\right)+\varepsilon^2H_2^{\varepsilon}(r).
   \end{eqnarray}
We note that in the case where the fluid flow is generated due to $N$ point forces, the linearity of the Brinkman equation allows the resulting flow to be written as
             \begin{eqnarray}
\mu{\bf u}(\bf{x})=\sum_{k=1}^NM_{\varepsilon}({\bf x}-{\bf x}_k){\bf f}_{k},\label{VelFluid}
   \end{eqnarray}
where $k=1, \cdots, N$ and $M_{\varepsilon}({\bf \hat{x}}_k)=H_1{\bf I}+{\bf \hat{x}}_k{\bf \hat{x}}_kH_2$ for ${\bf \hat{x}}_k={\bf x}-{\bf x}_k$ and identity matrix ${\bf I}$. Note that ${\bf x}=(x,y,z)$ is a point in the fluid and force ${\bf f}_k$ is located at ${\bf x}_k$. Eq.~(\ref{VelFluid}) is compactly written from Eq.~(\ref{RegVelBrink}) and determines the velocity field on the fluid domain at any given point ${\bf x}$. Explicitly, ${\bf f}_k=(f_k^x,f_k^y,f_k^z)$ where the force components are the forces in the $x, y$ and $z$ directions, respectively.

\subsection{ Test Cases}
For all test cases, the number of discretization points $N$ depends on the length $L$ of the swimmer and are evenly spaced with $\Delta s=0.01$. 
Unless otherwise stated, the regularization parameter $\varepsilon$ is 0.01.

\subsubsection{Planar Bending}
We first compare the numerical data obtained from the MRB with the asymptotic swimming speed for the case of planar bending. Consider an undulating filament parametrized by the following space curve equation as 
     \begin{eqnarray}
x(s,t)=s, \hspace{5mm}y(s,t)=b\sin(ks - \omega t), \hspace{5mm} z(s,t)=0,
   \end{eqnarray}
       \begin{figure}[!htb]
\centering
\includegraphics*[width=.45\textwidth]{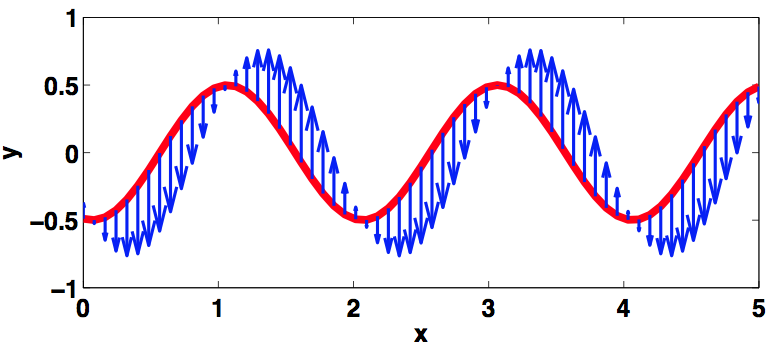}
\includegraphics*[width=.5\textwidth]{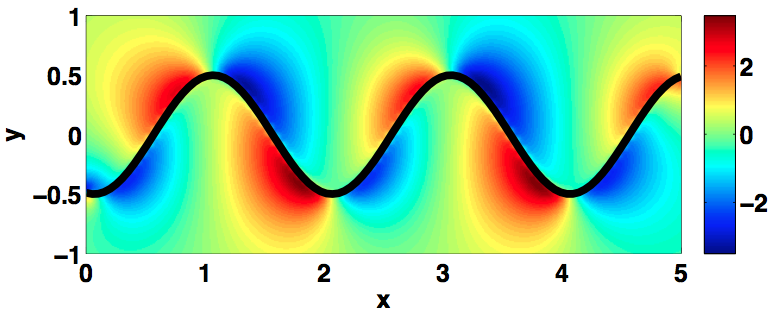}\\
{\bf (a)} \hspace{16em} {\bf (b)}
\caption{\cs{Numerical results shown in the $x$-$y$ plane for an undulating filament in a Brinkman fluid with $\gamma=10$, $b=0.5$, $L=5$, $\lambda=2$, and $\omega=2\pi$. (a) The velocity field is shown along the structure. (b) The corresponding pressure map of the fluid domain. }} \label{planarswimmer}
\end{figure}
   for $0<s<L$ where $s$ is a parameter initialized as arclength. The wavenumber is $k=2\pi/\lambda$ for wavelength $\lambda$, the bending amplitude is $b$, and $\omega$ is the constant angular speed. 
   At any given time $t\geq0$, the velocity of the flagellum is calculated by
        \begin{eqnarray}
u_x(s,t)=0, \hspace{5mm} u_y(s,t)=-b\omega \cos(ks - \omega t), \hspace{5mm} u_z(s,t)=0,
   \end{eqnarray}
 where $u_x, u_y$ and $u_z$ are the velocity components of $x, y$ and $z$, respectively. Fig.~\ref{planarswimmer}(a) shows the sinusoidal swimmer with the velocity fields along the length of the swimmer in the $x$-$y$ plane.  The total velocity includes the velocity from the sinusoidal wave ${\bf u}_s({\bf x})=(u_x, u_y, u_z)$, the translation ${\bf U}_0=(U_0^x, U_0^y,U_0^z)$, and the rotation of the filament $\boldsymbol{\Omega}_0=(\Omega_0^x,\Omega_0^y,\Omega_0^z)$ as: 
\begin{eqnarray}
\boldsymbol{ \mathcal{V}}={\bf u}_s({\bf x})+{\bf U}_0+\boldsymbol{\Omega}_0\times{\bf x}_k,\label{totalvel}
   \end{eqnarray}
where $\boldsymbol{ \mathcal{V}}$ is defined similarly to Eq.~(\ref{VelFluid}) and for simplicity, we choose $\mu=1$. Unless specified, the superscripts in translational and rotational velocity components are of the $x, y$ and $z$ components, not the partial derivatives. We note that ${\bf f}_k$, ${\bf U}_0$ and $\boldsymbol{\Omega}_0$ are constants at each time point which can be found by coupling Eq.~(\ref{totalvel}) with the force-free and torque-free conditions. That is,
           \begin{eqnarray}
\boldsymbol{ \mathcal{V}}-{\bf U}_0-\boldsymbol{\Omega}_0\times{\bf x}_k&=&{\bf u}_s({\bf x}_k),\label{totalvel1}\\
\sum_{k=1}^N{\bf f}_k&=&0,\label{forcefree}\\
\sum_{k=1}^N{\bf f}_k\times {\bf x}_k&=&0.\label{torquefree}
   \end{eqnarray}
In Eq.~(\ref{totalvel1}), for each value of $k$, $M_{\varepsilon}$ is a $3N\times3N$ matrix while the coefficients for ${\bf U_0}$ and $\boldsymbol{\Omega}_0$ will form $3\times3N$ matrices. The coefficient matrices in Eq.~(\ref{forcefree}) and Eq.~(\ref{torquefree}) are $(3N+6)\times3$. To determine ${\bf U_0}$, $\boldsymbol{\Omega}_0$, and $\bf{f}$, we solve Eq.~\eqref{totalvel1}-\eqref{torquefree}. We can then compute pressure using the regularized version of Eq.~\eqref{pBrinkman}. In Fig.~\ref{planarswimmer}(b), the pressure in the $x$-$y$ plane is shown where we note larger variations in pressure close to the swimmer.

The numerical results for the translational velocity ${\bf U}_0$ will be used to compare to the asymptotic swimming speed $U_{\infty}$ derived in Eq.~(\ref{velasymexpand}). The results are presented in Fig.~\ref{AsymNumer}.
       \begin{figure}[!htb]
\centering
\includegraphics*[width=.45\textwidth]{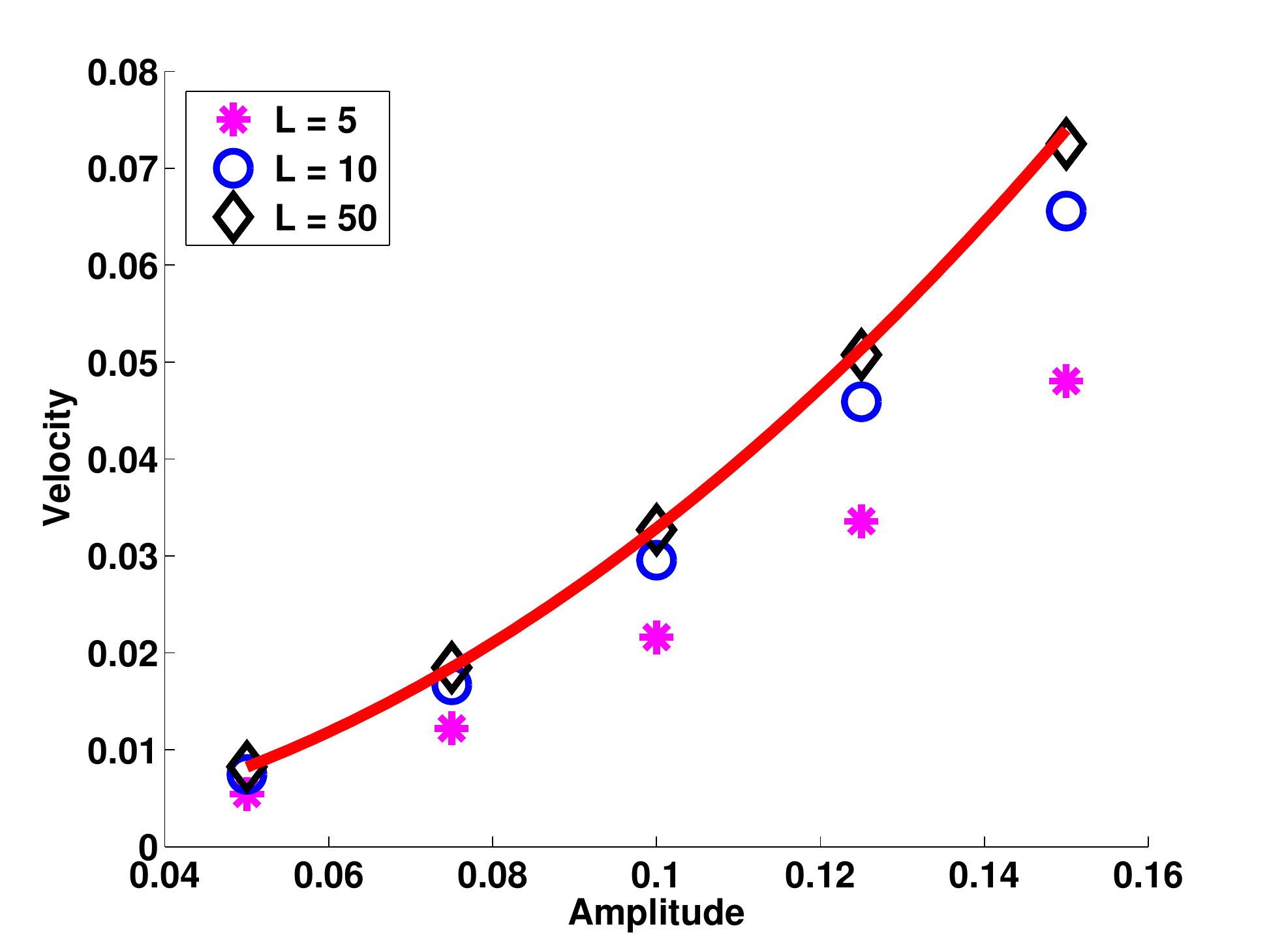}
\includegraphics*[width=.45\textwidth]{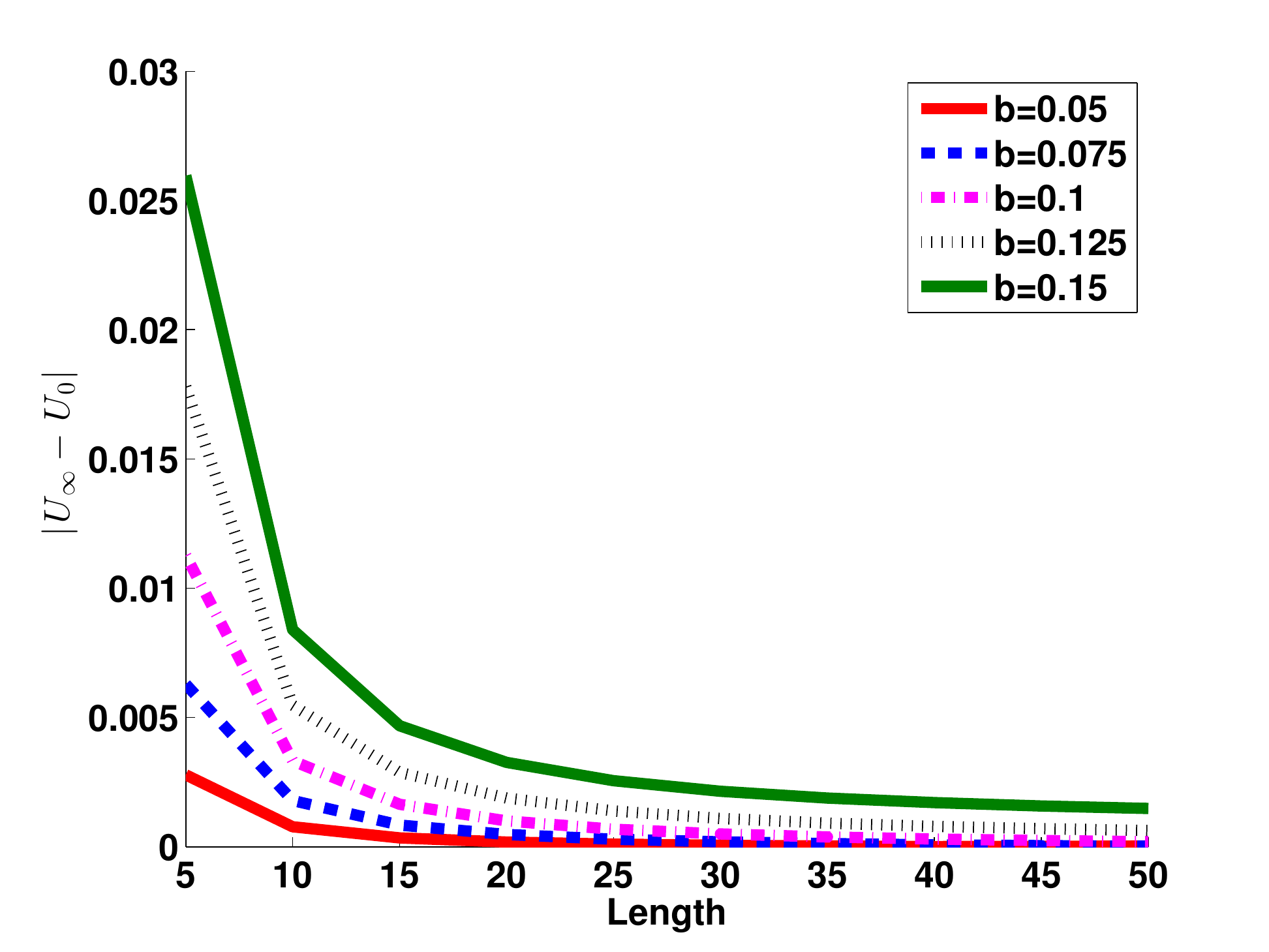}\\
{\bf (a)} \hspace{16em} {\bf (b)}
\caption{\cs{ (a) The comparison between the asymptotic swimming speed (solid line) with the numerical data (marker points) for different amplitudes when the length of the swimmers are $L=5$, $L=10$, and $L=50$ and the permeability is $\gamma=1$. (b) The difference between the asymptotic values with the numerical results for different amplitudes, and different lengths for $\gamma=1$. The wavelength is taken to be $\lambda=5$ for all simulations.}}\label{AsymNumer}
\end{figure}
Hereinafter, the wavelength is taken to be $\lambda=5$, $t=2\pi$ and $\omega=2\pi$. We prescribe five different amplitudes for the simulations as $b=0.05,~ 0.075,~ 0.1,~ 0.125$ and $0.15$. We set the permeability $\gamma=1$ and study the effects of swimmer length on swimming speed. We observe in Fig~\ref{AsymNumer}(a) that the numerical data (marker points) have good agreement with the asymptotic analysis (solid line) with a longer length ($L=50$). We also plot the difference between the asymptotic values with the numerics for different lengths and different amplitudes when $\gamma=1$ as shown in Fig.~\ref{AsymNumer}(b). As the length increases, the differences decrease. This shows that finite-length swimmers will swim \emph{slower} than the asymptotic predictions and this difference decreases for smaller amplitude $b$ (with fixed $\omega$ and $\lambda$). We also observe the same trend for different permeabilities $\gamma=0.1$ and $\gamma=10$ in Fig.~\ref{DifferentGamma}(a)-(b). We note that when looking closely at the error in Fig.~\ref{AsymNumer}(b) for $\gamma=1$ and Fig.~\ref{DifferentGamma}(a) for $\gamma=0.1$, the error is slightly larger for smaller permeability. Thus, the infinite-length cylinder swimming speed captures the swimming speed of a finite-length swimmer with more accuracy for larger permeability.
       \begin{figure}[!htb]
\centering
\includegraphics*[width=.4\textwidth]{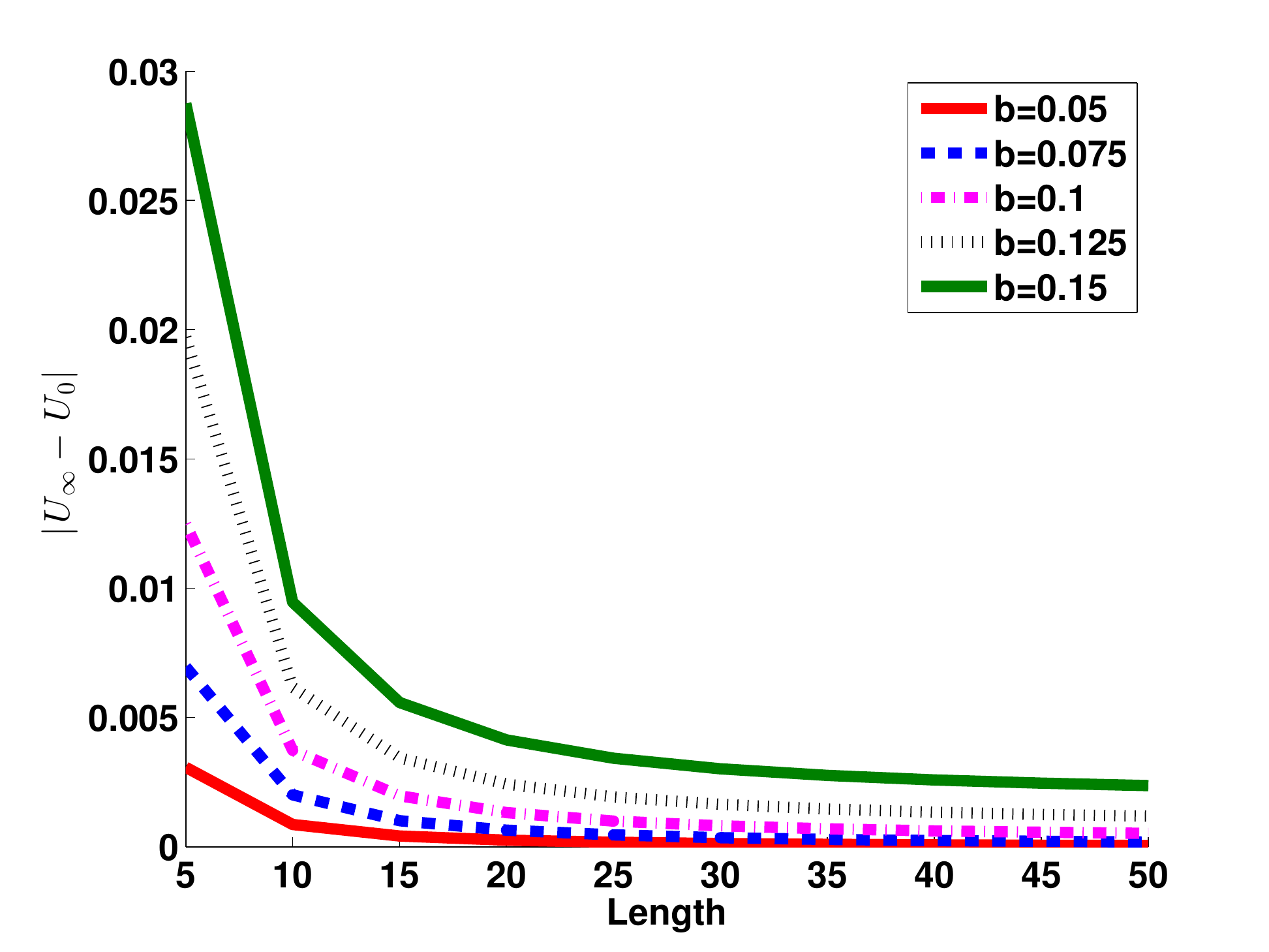}
\includegraphics*[width=.4\textwidth]{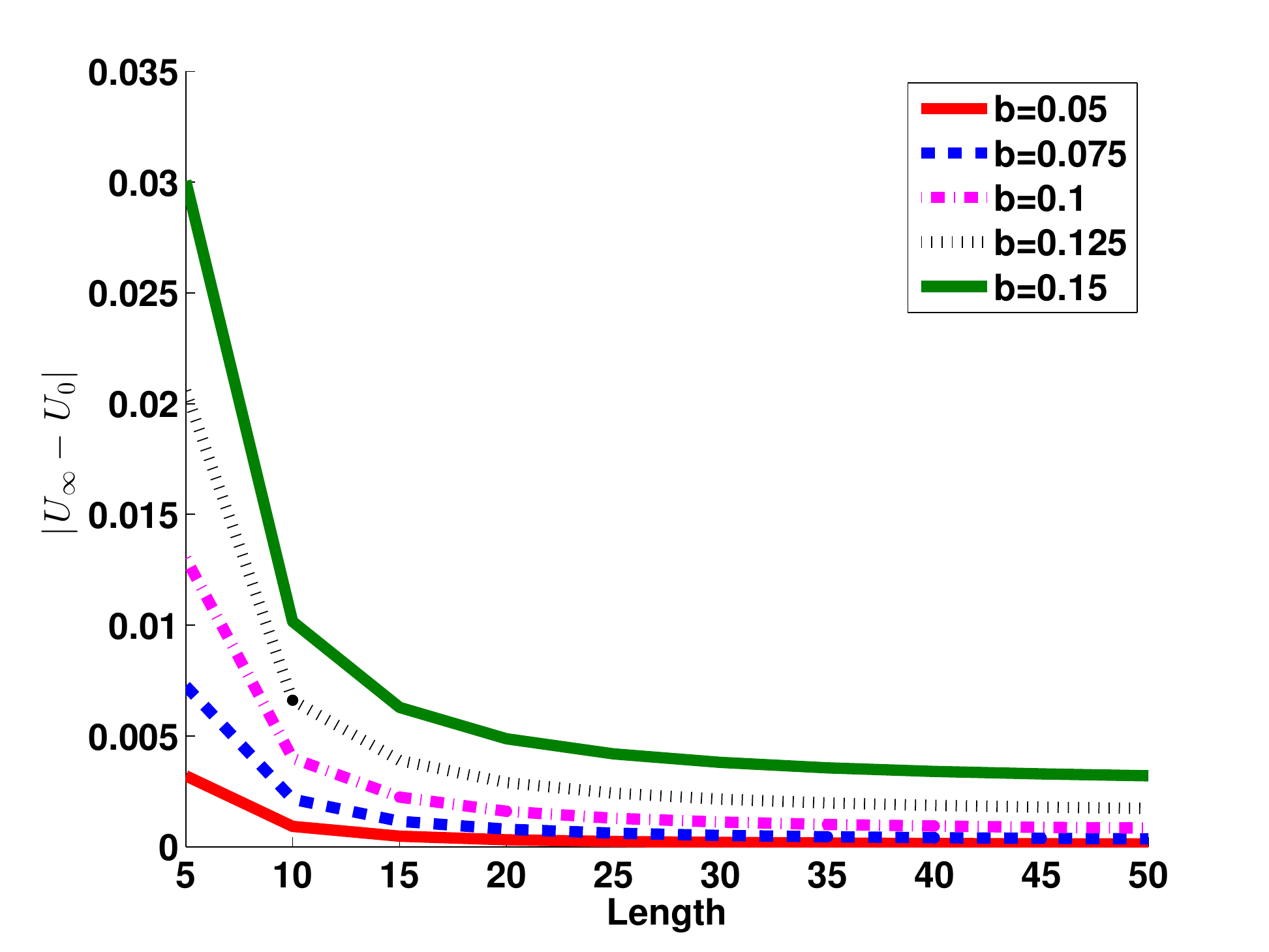}\\
{\bf (a)} \hspace{13em} {\bf (b)}

\caption{\cs{Difference between analytical results and numerical data for different amplitudes and different lengths when (a) $\gamma=0.1$ and (b) $\gamma=10$.
}} \label{DifferentGamma}
\end{figure}

In addition to the translational velocity, we can calculate the angular velocity for different parameters. We have five different amplitudes for the swimmer ranging from $b=0.05$ to $b=0.15$ with five different permeabilities $\gamma=0.01,~ 0.1,~ 1,~ 10$ and $\gamma=100$.    
       \begin{figure}[!htb]
\centering
\includegraphics*[width=.32\textwidth]{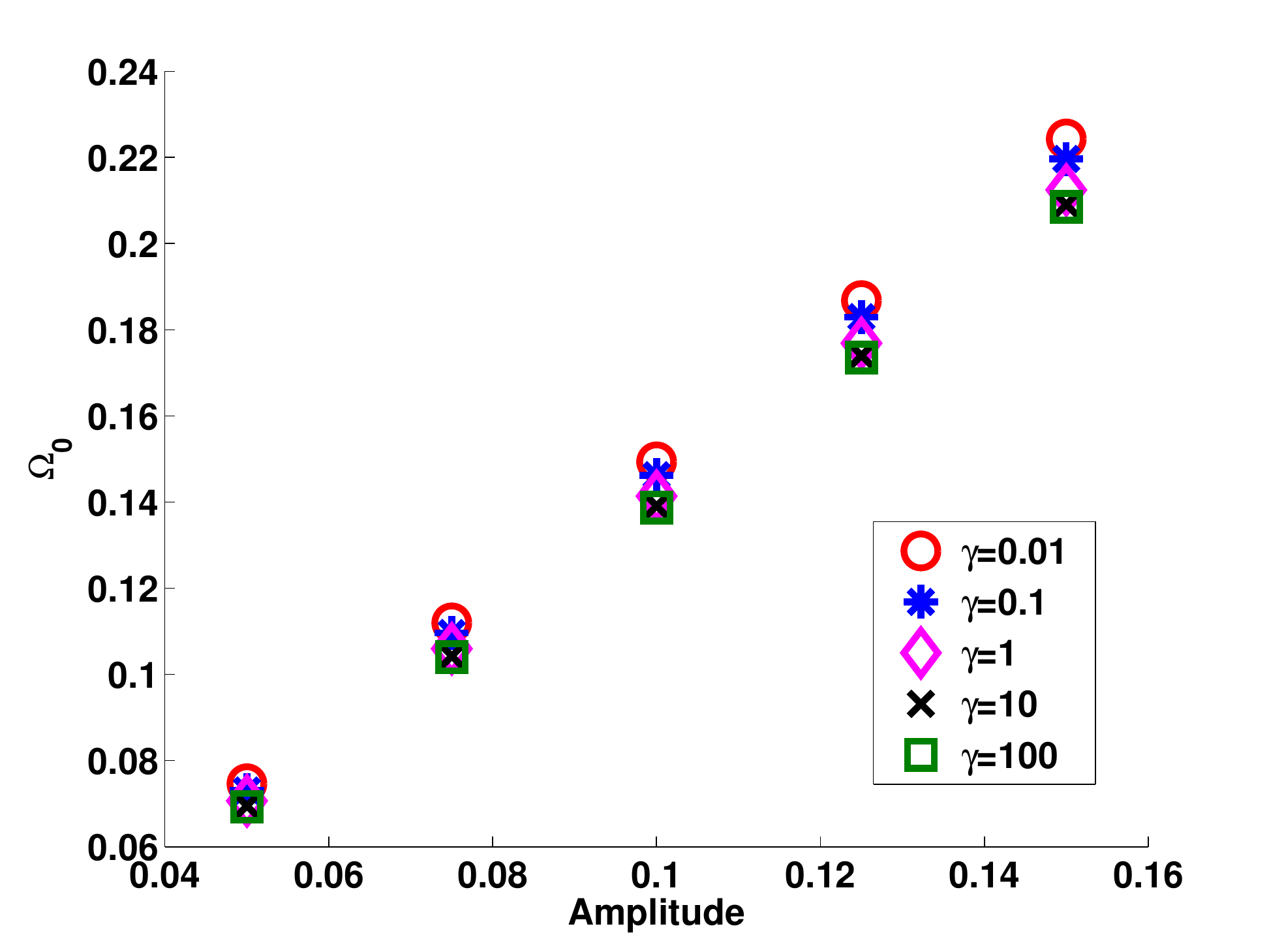}
\includegraphics*[width=.32\textwidth]{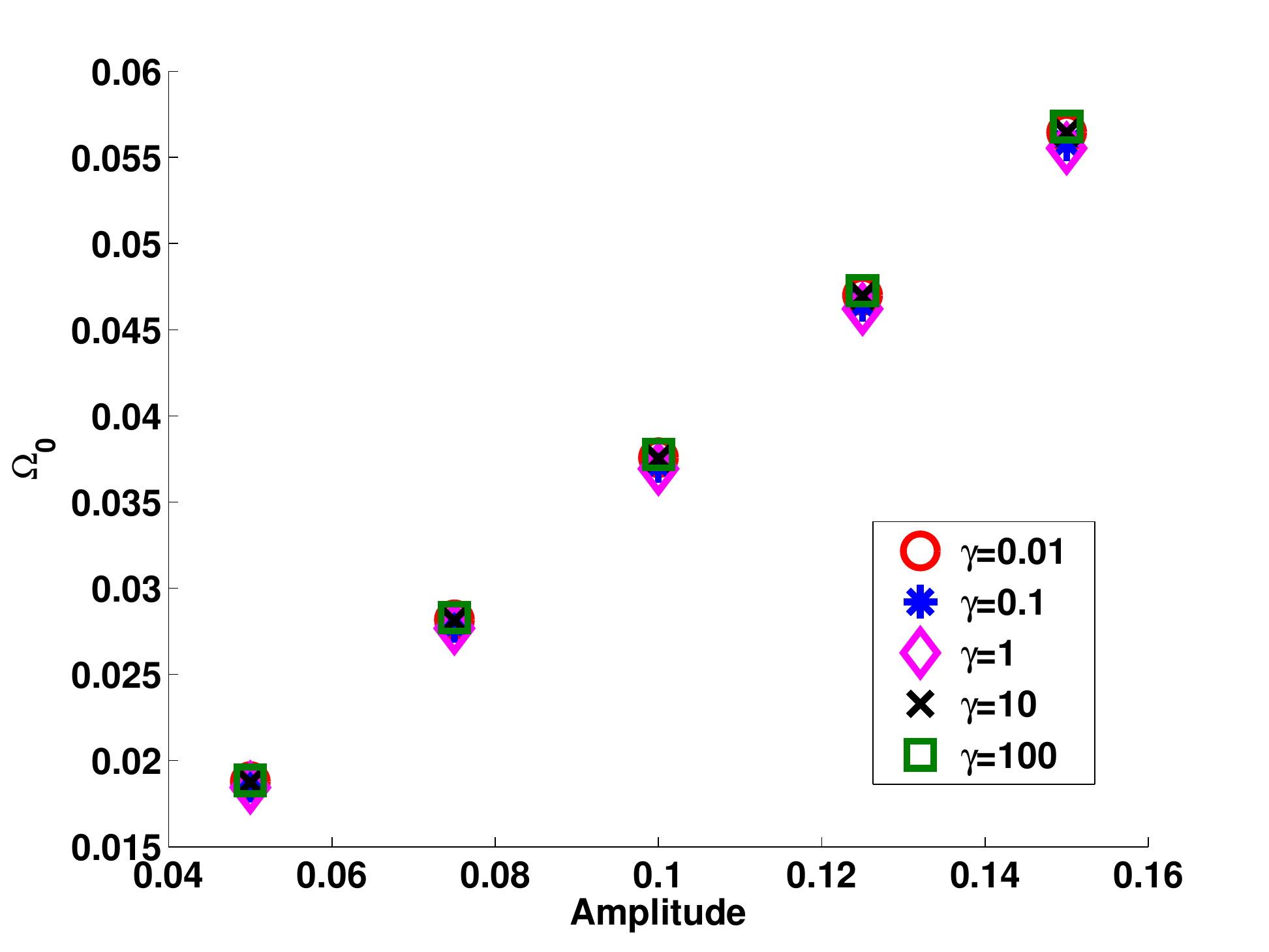}
\includegraphics*[width=.32\textwidth]{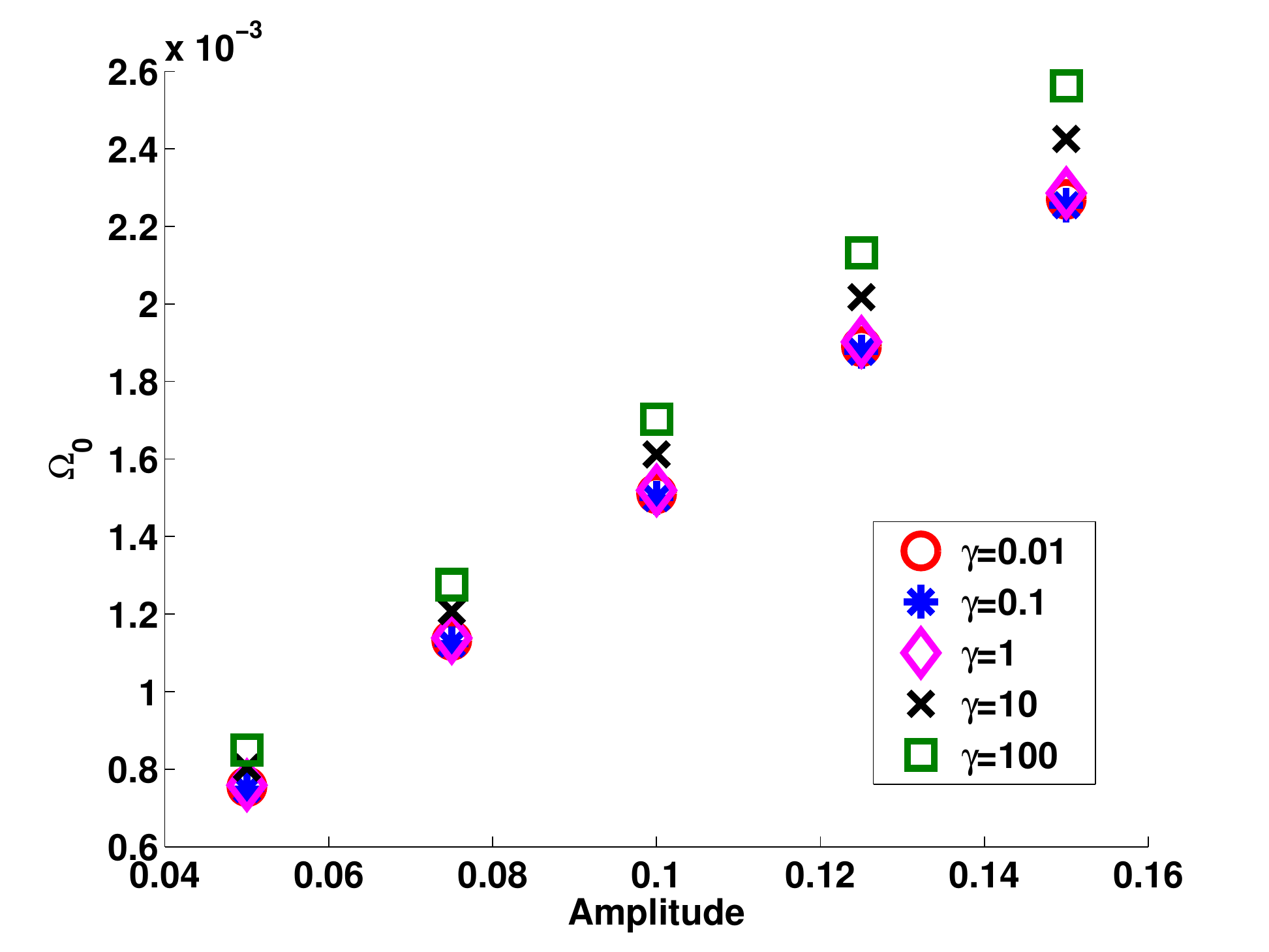}\\
\hspace{0.5em}{\bf (a)} \hspace{12em} {\bf (b)}\hspace{12em} {\bf (c)}
\caption{\cs{Angular velocity for swimmers for five different permeabilities $\gamma=0.01,~ 0.1,~ 1,~ 10, ~100$ with different amplitudes $b= 0.05,~ 0.075, ~0.1,~ 0.125,~ 0.15$. (a) $L=5$, (b) $L=10$, (c) $L=50$.}} \label{angularplots}
\end{figure}
Fig.~\ref{angularplots}(a) shows that the angular velocity when $L=5$ increases linearly as the amplitude increases. We capture the same behavior for longer swimmers (at $L=10$ in Fig.~\ref{angularplots}(b) and $L=50$ in Fig.~\ref{angularplots}(c)). 
We note that the angular velocity is much larger in the case of small length; in order for a swimmer of shorter length to achieve a prescribed amplitude, the angular velocity increases.} 

\cs{\subsubsection{Helical Bending Waves}
For this test case, we calculate the external torque exerted on the filament by the surrounding fluid. Consider the right-handed helix where the configuration is parameterized by the 3D space curve as 
    \begin{figure}[!htb]
\centering
\includegraphics*[width=.3\textwidth]{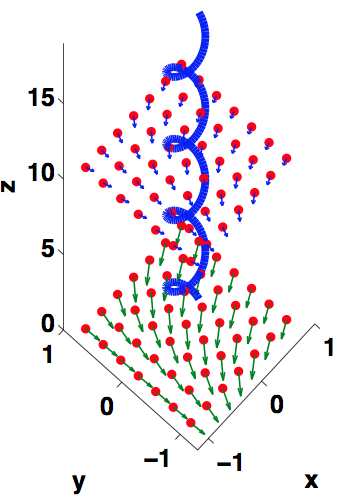}
\caption{\cs{Right-handed helix immersed in a Brinkman fluid with, $r_1=0.25$, $L=20$ and $\lambda=5$. The flow field is shown at $z=0$ and $z=10$. }} 
\end{figure}
 \begin{eqnarray}
x(s,t)=r_1\cos(ks+\omega t),\hspace{5mm} y(s,t)=r_1\sin(ks+\omega t), \hspace{5mm} z(s,t)=r_2s+\mathcal{U}t,
   \end{eqnarray}
for $s, k, \omega$ defined as above, $r_1$ is the radius of the helix (or the amplitude), $r_2$ is a constant defined as $r_2=\cos\theta=\sqrt{1-k^2r_1^2}$ where $\theta$ is the pitch angle, 
and $\mathcal{U}$ is the constant propulsion velocity. The prescribed helical configuration gives the velocity of the helix as 
 \begin{eqnarray}
u(s,t)=-r_1\omega\sin(ks+\omega t),\hspace{5mm} v(s,t)=r_1\omega\cos(ks+\omega t), \hspace{5mm} w(s,t)=\mathcal{U}.
   \end{eqnarray}
The torque is calculated as \cite{JungFauci07}
           \begin{eqnarray}
           {\bf \mathcal{T}}=\int_{\Gamma}{\bf g}_k\times{\bf x}_k~ds,\label{TorqueVal}
   \end{eqnarray}
  where $\Gamma$ is the helix (centerline of the flagella) and ${\bf g}_k$ is the surface force (traction) applied on the filament. 
 The torque is then numerically approximated by
              \begin{eqnarray}
           {\bf \mathcal{T}}=\sum_{k=1}^N\left({\bf g}_k\times{\bf x}_k\right)\Delta s,\label{TorqueDisc}
   \end{eqnarray}
   which we will compare with the analytical solution $T_{\infty}$ in (\ref{torqueanalysis}).
    \begin{figure}[!htb]
\centering
\includegraphics*[width=.4\textwidth]{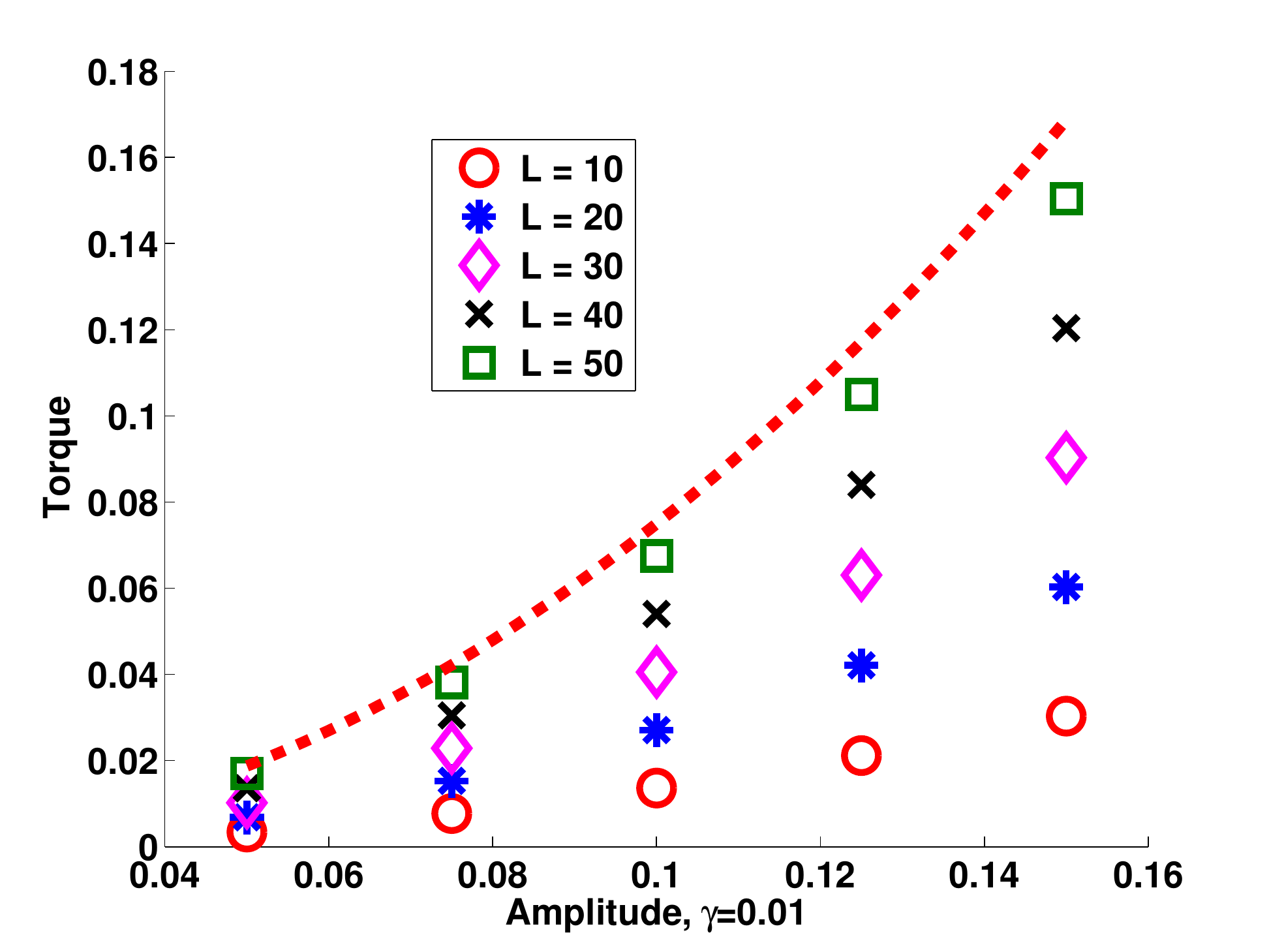}
\includegraphics*[width=.4\textwidth]{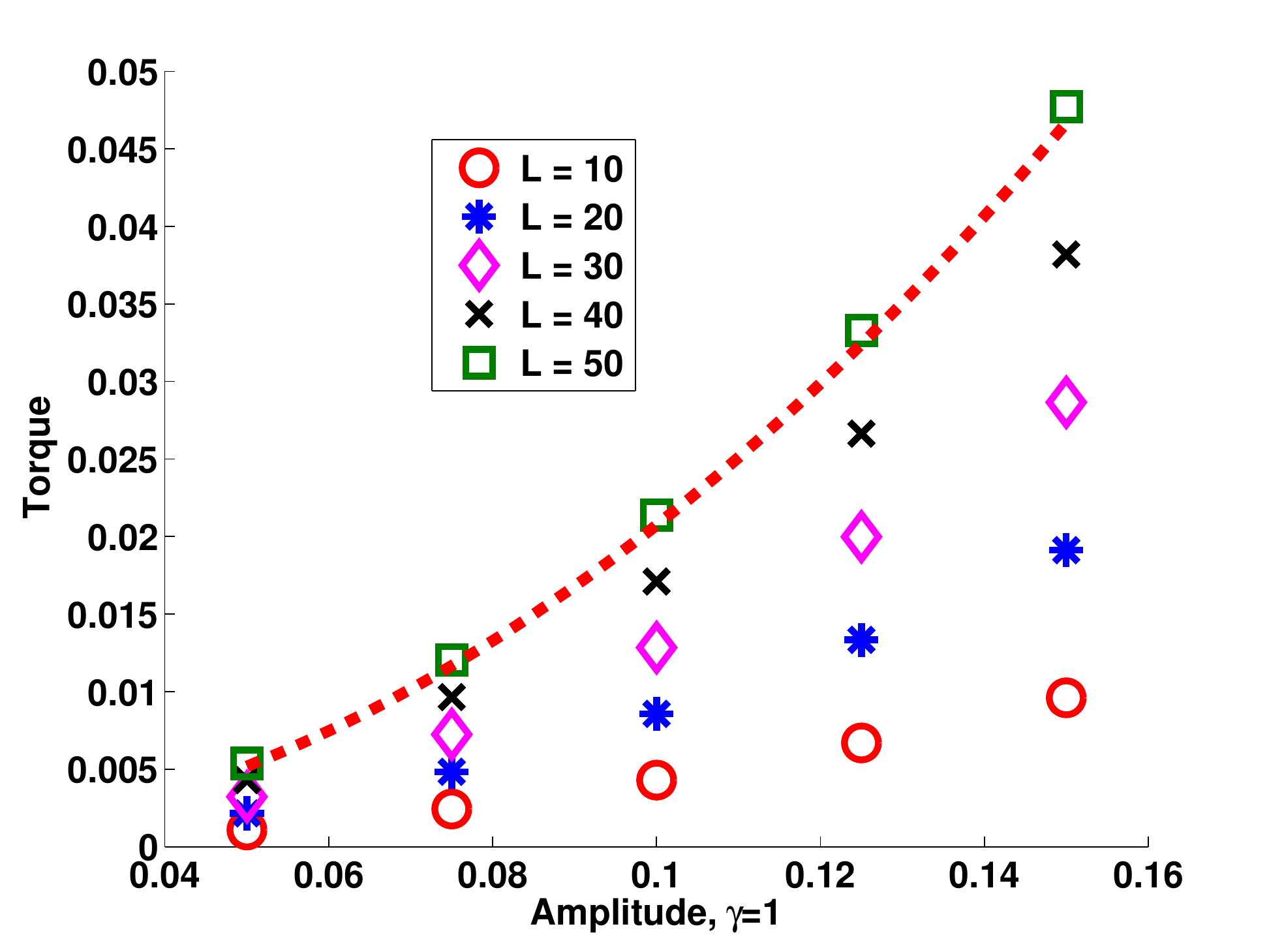}\\
{\bf (a)} \hspace{16em} {\bf (b)}\\
\includegraphics*[width=.4\textwidth]{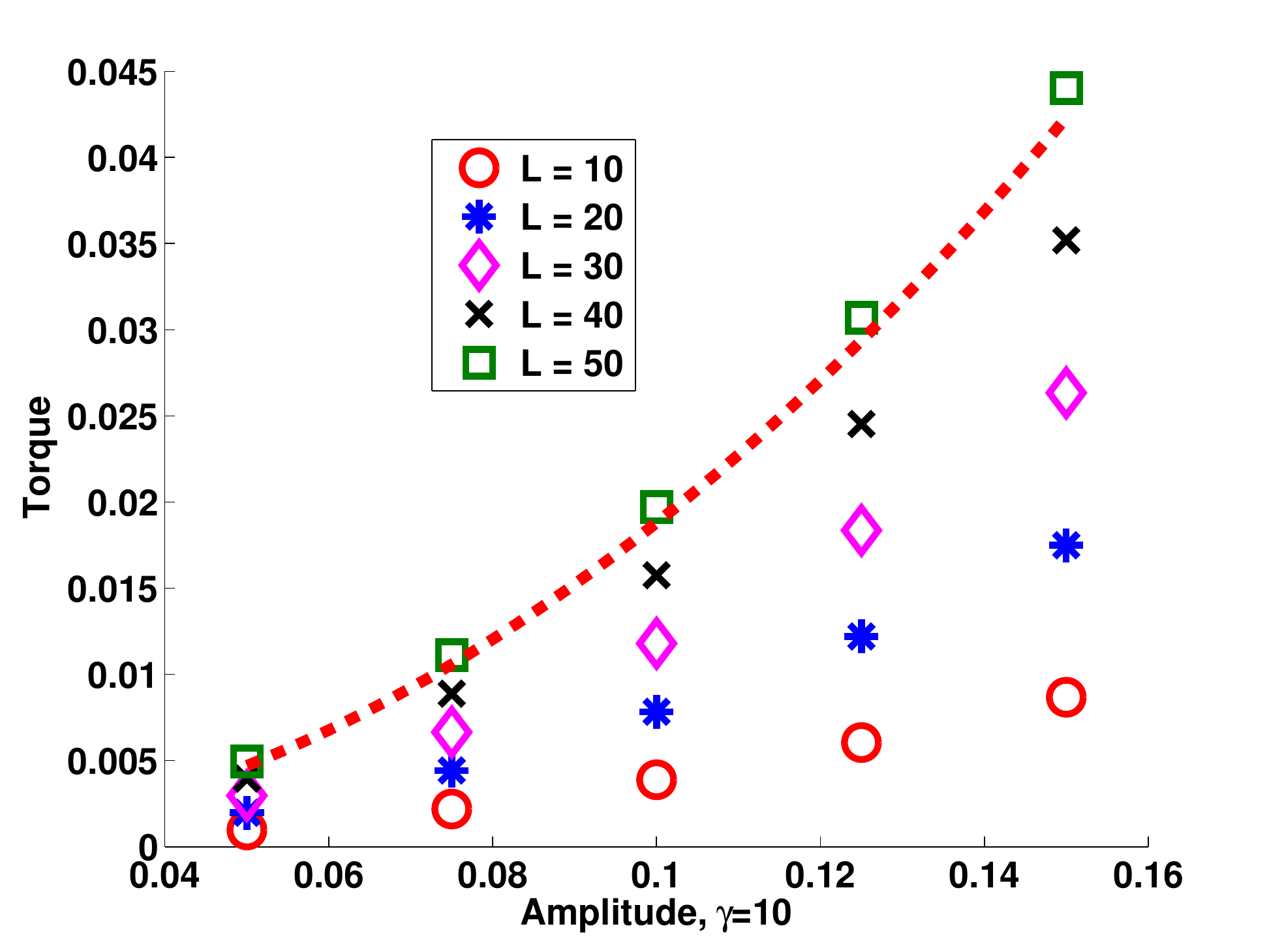}
\includegraphics*[width=.4\textwidth]{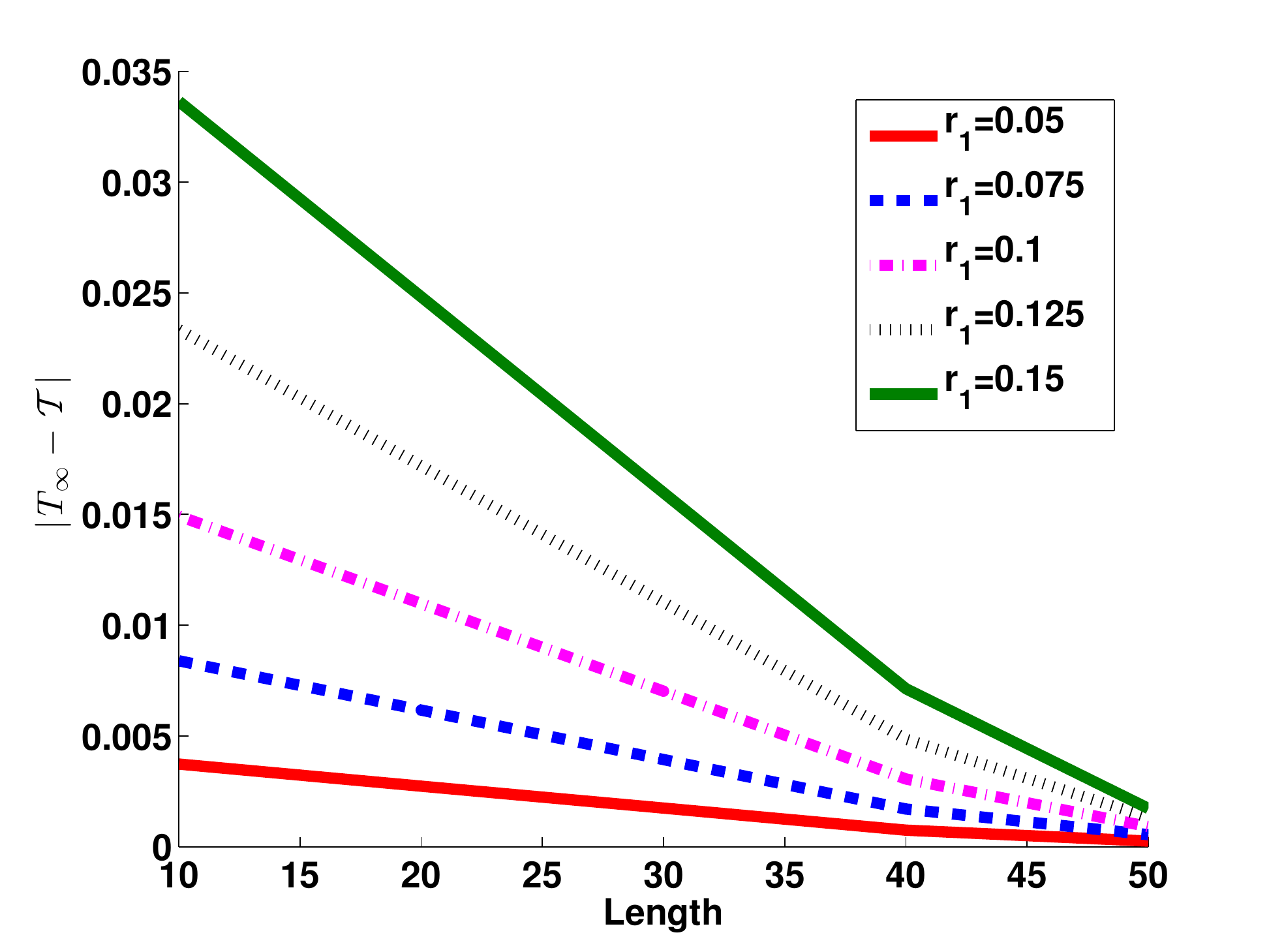}\\
{\bf (c)} \hspace{16em} {\bf (d)}
\caption{\cs{The comparison between the asymptotic velocity (dash-line) with the numerical data (marker points) for different amplitudes and different length, $L=10,~20,~30,~40$ and $L=50$ with $a=0.05$ when (a) $\gamma=0.01$ and $\varepsilon=0.01$, (b)  $\gamma=1$ and $\varepsilon=0.0055$, and (c) $\gamma=10$ and $\varepsilon=0.005$. (d) The difference between the asymptotic values with the numerical results for different amplitudes, and different lengths for $\gamma=10$.}}\label{FigTorqueNew}
\end{figure}

In Fig.~\ref{FigTorqueNew}, we present results for the helical bending case for amplitudes ($r_1=b$) in the range of 0.05 to 0.15 and set the wavelength at $\lambda=5$. For a filament of radius $a=0.05$ and permeability $\gamma=0.01$, we observe an increase in torque as amplitude increases in both the asymptotic analysis and the computational solutions. 
We note that similar to the swimming speeds for the case of planar bending, the asymptotics greatly \textit{overestimate} the torque for shorter length filaments. Similarly, for the cases of $\gamma=1$ and $\gamma=10$ (shown in Fig.~\ref{FigTorqueNew}(b)-(c)), the analytical results for an infinite-length spiral cylinder overestimate the torque for the finite-length spiral filament. 
The difference between the numerical data and analytical solutions is also plotted in Fig.~\ref{FigTorqueNew}(d) for $\gamma=10$ to show that for each fixed amplitude $r_1$, as the length increases, the difference decreases. We note that previous computational studies using the MRB have observed that the optimal numerical regularization parameter $\varepsilon$ varies for each $\gamma$ and can be sensitive for torque calculations \cite{Cortez10}. We also observed this sensitivity and decreased the regularization parameter as permeability increased to report the best fit with the asymptotics at longer lengths (values given in Figure caption). }

\section{Discussion}
In this paper, we have analyzed an infinite-length cylinder undergoing periodic bending in a fluid governed by the Brinkman equation, which is a model for flow through a porous medium. Motivated by organisms that exhibit undulatory locomotion such as spermatozoa, we focus on the case where the radius of the cylinder is small in comparison to the fiber spacing. We find that propulsion in the case of planar and spiral bending is enhanced with larger fluid resistance for specific combinations of wavenumber, cylinder thickness, and permeability. \cs{In Section \ref{rangeassumption}, we show that for a sufficiently small volume fraction of fibers, mammalian spermatozoa will observe an enhancement in swimming speed. Our calculations show that the  mesh spacing is several orders of magnitude larger than the cylinder thickness, allowing room for the swimmer to navigate between stationary fibers. This analysis provides insight into the sperm thickness and wavelength that we observe in nature; perhaps they have been optimized to provide enhanced swimming speeds in oviductal fluids. }

The observed enhancement in swimming speed for the infinite-length cylinder with planar bending is similar to the case for a swimming sheet in a Brinkman fluid \cite{Leshansky09}. \cs{We note that in both the 2D and 3D cases, as the resistance is reduced to zero, the corresponding swimming speeds in a Stokes fluid are recovered. For a fixed amplitude, wavenumber, and cylinder thickness, the ratio between the swimming speeds of the infinite-length  cylinder and sheet is approximately 0.8 in a Stokes fluid (using Eq.~\eqref{Stratio}). We observe that the ratio of the asymptotic swimming speeds of the 3D infinite-length cylinder and 2D sheet in a Brinkman fluid vary greatly as the scaled resistance increases, decreasing from 0.8 to 0.1. Thus, as the scaled resistance increases, potential rotational effects may play a large role in decreasing the swimming speed of the 3D infinite-length cylinder (in comparison to the 2D sheet). This highlights the importance of the analysis presented here for the 3D infinite-length swimmer, especially when trying to understand the role of larger resistance on swimming speeds. }

Relative to the Stokes case, the infinite-length sheet and cylinder swim slower in a viscoelastic fluid \cite{Fu07,Fu09,Lauga07}. However, the results reported here for an infinite-length cylinder and previous work for the sheet \cite{Leshansky09} show that added fluid resistance enhances swimming speeds relative to the Stokes case. Thus, a potential enhancement in swimming speed can be observed when a low volume fraction of obstructions do not have a frequency dependent response or when the polymer relaxation time is fast. In contrast to a viscoelastic fluid, the Brinkman fluid model assumes that the fibers or polymers in the fluid are stationary.

Through a detailed mathematical analysis, we have derived the work for planar waves. We observed that larger resistance (smaller permeability) results in a large increase in work. \cs{This increase in work will occur when the cylinder thickness satisfies the inequality in Eq.~\eqref{zetaineq}. We note that in the asymptotic derivation, we have assumed prescribed kinematics. Thus, as the resistance increases, it requires more work to maintain planar bending with the same amplitude and wavenumber.} When building artificial microswimmers in a porous medium, one must consider the amount of energy required to have the swimmer bend \cite{Sanders09}. This could be a constraint on reaching higher swimming speeds in fluids with larger resistance.
 
\cs{We have compared our asymptotic solutions to computations of finite-length swimmers with prescribed kinematics using the method of regularized Brinkmanlets. For cylinders of sufficient length, the asymptotic swimming speeds match well with the computations. We note that the asymptotic analysis is able to capture the trends of swimming speed in terms of the dependence on permeability and amplitude. However, it \textit{overestimates} the swimming speed for shorter length filaments. This is important to consider when using asymptotic swimming speeds to make predictions of the behavior of finite-length swimmers. Additionally, we have observed that the analytical results \textit{overestimate} the torque for a finite-length filament with helical propagating wave.\\
\indent For prescribed kinematics, we note that the asymptotic and computational swimming speeds calculated  increase as amplitude and resistance increase. The asymptotic analysis presented here provides the swimming speed given that a swimmer could attain the given prescribed kinematics in a fluid with permeability $\gamma$.
Previous studies have observed non-monotonic changes in swimming speed for finite-length swimmers with increasing fluid resistance for planar swimmers, where the achieved amplitude of bending is an emergent property of the fluid-structure interaction \cite{Cortez10,Olson15a}. In these studies, finite-length swimmers were not able to achieve large amplitude bending as the permeability is decreased. In addition, experimental studies have shown that the emergent waveforms and swimming speeds will depend strongly on the fluid environment \cite{Smith09b,Suarez92}.  Thus, it is important to put the asymptotic results in the context of finite-length swimmers where certain ranges of bending kinematics are not observed in gels or fluids with small volume fractions of fibers.  \\
\indent In this computational study of a finite-length filament undergoing periodic lateral bending, we observed a large increase in angular velocity as the swimmer length decreases. Additionally, angular velocity increased linearly as amplitude increased for a fixed beat frequency. Sperm cells have been observed to `roll' as they swim (simultaneous rotation of the sperm cell body and flagellum) \cite{Babcock14,Smith09b}. Specifically, human sperm were found to increase rolling from ~1.5 Hz to 10 Hz and decrease amplitude as the viscosity of methylcellulose solutions was decreased \cite{Smith09b}. 
In our computational study, angular rotation (rolling) varies linearly with amplitude and is much smaller than the experimental data. However, we are not accounting for the dynamics of a cell body and have prescribed kinematics. 
It will be interesting to study three-dimensional computational models of finite-length swimmers with cell bodies and emergent kinematics in the future to fully understand swimming speed and angular velocity as a function of the permeability.} 

\section{Appendix}
\subsection{Derivation of surface cylinder location for planar bending waves}\label{Rderive}

The surface points of the cylinder are defined by Eq.~\eqref{rEq}. Using Fig.~\ref{CylGeomPic}, 
we observe that $\theta'=\theta+\xi$ and use this to rewrite the equation as:
\begin{eqnarray*}
r^2&=&a^2+b^2\sin^2s+2ab\sin s\cos(\theta+\xi)\\
&=&a^2+b^2\sin^2s+2ab\sin s(\cos\theta \cos\xi-\sin\theta \sin \xi)
\end{eqnarray*}
where $\sin \xi=\dfrac{b\sin s\sin\theta}{a}$. For small $\xi$, we have:
\begin{eqnarray*}
r^2&=&a^2+b^2\sin^2s+2ab\sin s\cos\theta-2b^2\sin^2s\sin^2\theta\\
&=&a^2+2ab\sin s\cos\theta+b^2\sin^2s(1-2\sin^2\theta)\\
&=&a^2+2ab\sin s\cos\theta+b^2\sin^2s\cos^2\theta-b^2\sin^2s\sin^2\theta\\
&=&(a+b\sin s\cos\theta)^2-b^2\sin^2s\sin^2\theta\\
&=&(a+b\sin s\cos\theta)^2\left[1-\dfrac{b^2\sin^2s\sin^2\theta}{(a+b\sin s\cos\theta)^2}\right]\\
r&=&(a+b\sin s\cos\theta)\sqrt{1-\dfrac{b^2\sin^2s\sin^2\theta}{(a+b\sin s\cos\theta)^2}}.
\end{eqnarray*}
We can then arrive at the final equation for $r$:
\begin{equation}
r=a\left[1+\dfrac{b}{a}\sin s\cos\theta+O\left(\dfrac{b}{a}\right)^2\right].
\end{equation}
\subsection{Asymptotics}\label{asymptoticexpand}
We wish to calculate the swimming speed of the infinite cylinder.  The {velocity components} are expanded up to the second order about $\zeta=\zeta_1=ka$:
   \begin{eqnarray}
   u&=&(u)_{\zeta=\zeta_1}+bk\cos\theta\sin s(u')_{\zeta=\zeta_1}+\cdots,\nonumber\\
      v&=&(v)_{\zeta=\zeta_1}+bk\cos\theta\sin s(v')_{\zeta=\zeta_1}+\cdots,\label{Taylorexpand}\\
         w&=&(w)_{\zeta=\zeta_1}+bk\cos\theta\sin s(w')_{\zeta=\zeta_1}+\cdots\nonumber,
        \end{eqnarray}
        where Eq.~\eqref{shapeBC} is used to rewrite $(\zeta-\zeta_1)$. Additionally, the velocity components $u$, $v$, and $w$ are expanded in the powers of $b/a$,
 \begin{equation}
 u=u^{(1)}+u^{(2)}+\cdots,\hspace{0.5cm}
 v=v^{(1)}+v^{(2)}+\cdots,\hspace{0.5cm}
 w=w^{(1)}+w^{(2)}+\cdots.\label{per_expand}
     \end{equation}
Substituting Eq.~\eqref{per_expand} into Eqs.~\eqref{Taylorexpand} and \eqref{vel3}:
   \begin{eqnarray}
   u&=&u_1\cos\theta\cos s+\left.u^{(2)}\right|_{\zeta=\zeta_1}+bk\cos\theta\sin s\cos \theta\cos s(u')_{\zeta=\zeta_1},\label{u1u2} \\
v&=&v_1\sin \theta \cos s +\left.v^{(2)}\right|_{\zeta=\zeta_1}+bk\cos\theta\sin s\sin \theta \cos s(v')_{\zeta=\zeta_1},\label{v1v2}\\
 w&=&w_1\cos \theta \sin s+\left.w^{(2)}\right|_{\zeta=\zeta_1}+bk\cos\theta\sin s\cos \theta \sin s(w')_{\zeta=\zeta_1}.\label{w1w2}
      \end{eqnarray}
By matching the above expansions with the boundary conditions in Eq.~\eqref{velBC}, we can determine the constant coefficients $A_{mn}$, $B_{mn}$ and $C_{mn}$ for each order of the expansion.

Specifically, the second order expansion is:
\begin{eqnarray}
u^{(2)}&=&-bk\cos\theta\sin s\cos \theta\cos s(u')_{\zeta=\zeta_1}=-\dfrac{1}{4}bk(u_1')_{\zeta=\zeta_1}(\sin 2s+\cos2\theta\sin 2s),\label{veluapp}\\
v^{(2)}&=&-bk\cos\theta\sin s\sin \theta \cos s(v')_{\zeta=\zeta_1}=-\dfrac{1}{4}bk(v'_1)_{\zeta=\zeta_1}\sin2\theta\sin2s,\label{velvapp}\\
w^{(2)}&=&-bk\cos\theta\sin s\cos \theta \sin s(w')_{\zeta=\zeta_1}=-\dfrac{1}{4}bk(w_1')_{\zeta=\zeta_1}(1-\cos2s+\cos2\theta-\cos2\theta\cos2s).\nonumber\label{velwapp}\\
   \end{eqnarray}
The coefficients of the velocity in the second order expansion can be evaluated as:
\begin{eqnarray*}
u_{02}=-\dfrac{1}{4}bk(u_1')_{\zeta=\zeta_1},\hspace{5mm} u_{22}=-\dfrac{1}{4}bk(u_1')_{\zeta=\zeta_1},\\
v_{22}=-\dfrac{1}{4}bk(v_1')_{\zeta=\zeta_1},\\
w_{02}=\dfrac{1}{4}bk(w_1')_{\zeta=\zeta_1},\hspace{5mm} w_{20}=-\dfrac{1}{4}bk(w_1')_{\zeta=\zeta_1}, \hspace{5mm} w_{22}=\dfrac{1}{4}bk(w_1')_{\zeta=\zeta_1}.
   \end{eqnarray*}
The cylinder will move at a speed of $U_{\infty}$ with respect to the fluid at infinity which is also the term that balances the constant expression in $w^{(2)}$. Then,
\begin{eqnarray*}
U_{\infty}=\dfrac{1}{4}bk(w_1')_{\zeta=\zeta_1=ka}.
   \end{eqnarray*}

\subsection{Derivation of surface cylinder location for spiral bending waves}\label{RderiveSpiral}
As shown in Fig. \ref{SpiralGeomPic}, the time-dependent contour caused by the spiral bending wave is defined as
\begin{eqnarray*}
a^2&=&r^2+b^2-2br\cos(\theta-s)\\
&=&[r-b\cos(\theta-s)]^2+b^2\sin^2(\theta-s),
\end{eqnarray*}
or equivalently,
\begin{eqnarray*}
r&=&b\cos(\theta-s)+\sqrt{a^2-b^2\sin^2(\theta-s)},\\
r&=&a\left[1+\dfrac{b}{a}\cos(\theta-s)+O\left(\dfrac{b}{a}\right)^2\right].
\end{eqnarray*}

\cs{\subsection{Analysis of the Asymptotic Functions}\label{AsymptoticFuncs}
We look more closely at the behavior of the velocity in Eq.~(\ref{velasymexpand}) and the work done in Eq.~(\ref{Wrad}). Rewriting in terms of the scaled resistance $\beta=\alpha/k$,
   \begin{eqnarray}
f(\beta)=\dfrac{U_{\infty}}{(1/2)b^2k^2U}&=&\dfrac{K_0(\zeta_1)-\dfrac{1}{2}\left(\dfrac{1}{\beta^2}+1\right)\log\left(1+\beta^2\right)}{K_0(\zeta_1)+\dfrac{1}{2}\left(\dfrac{1}{\beta^2}-1\right)\log\left(1+\beta^2\right)},\label{function_f}\\\nonumber\\
g(\beta)=\dfrac{\overline{W}}{\mu\pi b^2k^2U^2}&=&\dfrac{1}{K_0(\zeta_1)+\dfrac{1}{2}\left(\dfrac{1}{\beta^2}-1\right)\log\left(1+\beta^2\right)}\label{function_g}.
   \end{eqnarray}
The two functions are plotted in Fig.~\ref{VelWorkFunction1}. Using the condition in (\ref{K0equation}), $f(\beta)$ and $g(\beta)$ are positive functions and $f(\beta)$ is bounded by 1. The first derivatives of $f(\beta)$ and $g(\beta)$ with respect to $\beta$ are
   \begin{small}
      \begin{eqnarray}
f'(\beta)&=&\dfrac{2\beta^2\left\{K_0(\zeta_1)\left[-1+\log(1+\beta^2)+\dfrac{1}{\beta^2}\log(1+\beta^2)\right]-\dfrac{1}{2}\left(1+\dfrac{1}{\beta^2}\right)\log^2(1+\beta^2)\right\}}{(1+\beta^2)\left[K_0(\zeta_1)+\dfrac{1}{2}\left(\dfrac{1}{\beta^2}-1\right)\log\left(1+\beta^2\right)\right]^2},\\
g'(\beta)&=&\dfrac{(1+\beta^2)\log(1+\beta^2)+\beta^2(-1+\beta^4)}{(1+\beta^2)\left[K_0(\zeta_1)+\dfrac{1}{2}\left(\dfrac{1}{\beta^2}-1\right)\log\left(1+\beta^2\right)\right]^2}.
   \end{eqnarray}
      \end{small}
We observe that all terms in the denominator and the numerator of $g'(\beta)$ are always positive for all $\beta$ which implies $g(\beta)$ is an increasing function. On the other hand, the function inside the curly bracket of $f'(\beta)$ is positive when 
      \begin{eqnarray}
K_0(\zeta_1)>\dfrac{\left(1+1/\beta^2\right)\log^2(1+\beta^2)}{2[-1+\log(1+\beta^2)+(1/\beta^2)\log(1+\beta^2)]}\label{fxcondition}.
   \end{eqnarray}
 In other words, $f(\beta)$ is an increasing function when it satisfies the condition in (\ref{fxcondition}). 
    \begin{figure}[!htb]
\centering
\includegraphics*[width=.45\textwidth]{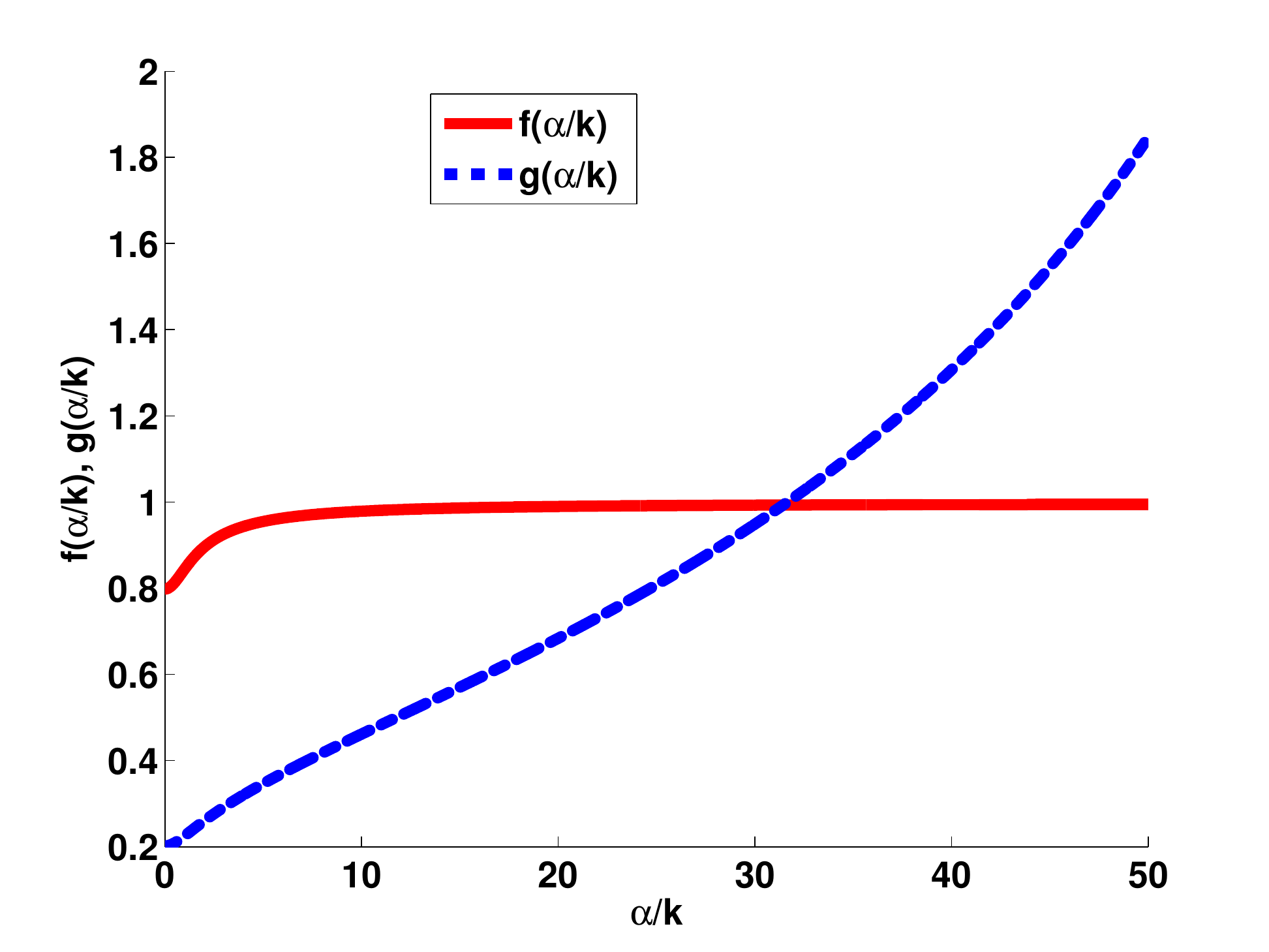}
\caption{\cs{The plot of the functions $f$ in Eq.~(\ref{function_f}) and $g$ in Eq.~(\ref{function_g}). }}
\label{VelWorkFunction1}
\end{figure}
We note that the expression $-1+\log(1+\beta^2)+(1/\beta^2)\log(1+\beta^2)$ is always positive. The Taylor expansions of $f(\beta)$ and $g(\beta)$ about $\beta\ll1$ are as follows:
       \begin{eqnarray}
f(\beta)\approx\dfrac{K_0(\zeta_1)-1/2}{K_0(\zeta_1)+1/2}+\mathcal{O}(\beta^2), \hspace{10mm}g(\beta)\approx\dfrac{1}{K_0(\zeta_1)+1/2}+\mathcal{O}(\beta^2).
   \end{eqnarray}
 This shows that when $\beta$ is small, $f(\beta)>g(\beta)$ as in Fig.~\ref{VelWorkFunction1}. When $\beta$ is large, we can expand the two functions in terms of the Puiseux series as: 
        \begin{eqnarray}
f(\beta)&\approx& 1+\dfrac{2\log(1/\beta)}{\beta^2\left[K_0(\zeta_1)+\log(1/\beta)\right]}+\mathcal{O}\left(\dfrac{1}{\beta^4}\right),\\
g(\beta)&\approx& \dfrac{1}{K_0(\zeta_1)+\log(1/\beta)}+\mathcal{O}\left(\dfrac{1}{\beta^2}\right).
   \end{eqnarray}
Clearly, $f(\beta)$ is bounded by 1 when $\beta$ is large while $g(\beta)$ is unbounded. The two formulations above give insight as to why a decrease in permeability $\gamma$ causes a small increase in swimming speed and a large increase on the rate of work done.

}

\section*{Acknowledgements} 
The work of N Ho and SD Olson was supported, in part, by NSF DMS 1413110. K Leiderman was supported, in part, by NSF DMS 1413078.\newpage

\begin{thebibliography}{51}%
\makeatletter
\providecommand \@ifxundefined [1]{%
 \@ifx{#1\undefined}
}%
\providecommand \@ifnum [1]{%
 \ifnum #1\expandafter \@firstoftwo
 \else \expandafter \@secondoftwo
 \fi
}%
\providecommand \@ifx [1]{%
 \ifx #1\expandafter \@firstoftwo
 \else \expandafter \@secondoftwo
 \fi
}%
\providecommand \natexlab [1]{#1}%
\providecommand \enquote  [1]{``#1''}%
\providecommand \bibnamefont  [1]{#1}%
\providecommand \bibfnamefont [1]{#1}%
\providecommand \citenamefont [1]{#1}%
\providecommand \href@noop [0]{\@secondoftwo}%
\providecommand \href [0]{\begingroup \@sanitize@url \@href}%
\providecommand \@href[1]{\@@startlink{#1}\@@href}%
\providecommand \@@href[1]{\endgroup#1\@@endlink}%
\providecommand \@sanitize@url [0]{\catcode `\\12\catcode `\$12\catcode
  `\&12\catcode `\#12\catcode `\^12\catcode `\_12\catcode `\%12\relax}%
\providecommand \@@startlink[1]{}%
\providecommand \@@endlink[0]{}%
\providecommand \url  [0]{\begingroup\@sanitize@url \@url }%
\providecommand \@url [1]{\endgroup\@href {#1}{\urlprefix }}%
\providecommand \urlprefix  [0]{URL }%
\providecommand \Eprint [0]{\href }%
\providecommand \doibase [0]{http://dx.doi.org/}%
\providecommand \selectlanguage [0]{\@gobble}%
\providecommand \bibinfo  [0]{\@secondoftwo}%
\providecommand \bibfield  [0]{\@secondoftwo}%
\providecommand \translation [1]{[#1]}%
\providecommand \BibitemOpen [0]{}%
\providecommand \bibitemStop [0]{}%
\providecommand \bibitemNoStop [0]{.\EOS\space}%
\providecommand \EOS [0]{\spacefactor3000\relax}%
\providecommand \BibitemShut  [1]{\csname bibitem#1\endcsname}%
\let\auto@bib@innerbib\@empty
\bibitem [{\citenamefont {Lauga}\ and\ \citenamefont {Powers}(2009)}]{Lauga09}%
  \BibitemOpen
  \bibfield  {author} {\bibinfo {author} {\bibfnamefont {E.}~\bibnamefont
  {Lauga}}\ and\ \bibinfo {author} {\bibfnamefont {T.}~\bibfnamefont {R.}~\bibnamefont {Powers}},\
  }\href@noop {} {\bibfield  {journal} {\bibinfo  {journal} {Rep Prog Phys}\
  }\textbf {\bibinfo {volume} {72}},\ \bibinfo {pages} {096601} (\bibinfo
  {year} {2009})}\BibitemShut {NoStop}%
\bibitem [{\citenamefont {Brennen}\ and\ \citenamefont
  {Winet}(1977)}]{Brennen77}%
  \BibitemOpen
  \bibfield  {author} {\bibinfo {author} {\bibfnamefont {C.}~\bibnamefont
  {Brennen}}\ and\ \bibinfo {author} {\bibfnamefont {H.}~\bibnamefont
  {Winet}},\ }\href@noop {} {\bibfield  {journal} {\bibinfo  {journal} {Ann Rev
  Fluid Mech}\ }\textbf {\bibinfo {volume} {9}},\ \bibinfo {pages} {339}
  (\bibinfo {year} {1977})}\BibitemShut {NoStop}%
\bibitem [{\citenamefont {Woolley}\ and\ \citenamefont
  {Vernon}(2001)}]{Woolley01}%
  \BibitemOpen
  \bibfield  {author} {\bibinfo {author} {\bibfnamefont {D.}~\bibnamefont
  {Woolley}}\ and\ \bibinfo {author} {\bibfnamefont {G.}~\bibnamefont
  {Vernon}},\ }\href@noop {} {\bibfield  {journal} {\bibinfo  {journal} {J Exp
  Biol}\ }\textbf {\bibinfo {volume} {204}},\ \bibinfo {pages} {1333} (\bibinfo
  {year} {2001})}\BibitemShut {NoStop}%
\bibitem [{\citenamefont {Smith}\ \emph
  {et~al.}(2009{\natexlab{a}})\citenamefont {Smith}, \citenamefont {Gaffney},
  \citenamefont {Blake},\ and\ \citenamefont {Kirkman-Brown}}]{Smith09Surf}%
  \BibitemOpen
  \bibfield  {author} {\bibinfo {author} {\bibfnamefont {D.}~\bibnamefont
  {Smith}}, \bibinfo {author} {\bibfnamefont {E.}~\bibnamefont {Gaffney}},
  \bibinfo {author} {\bibfnamefont {J.}~\bibnamefont {Blake}}, \ and\ \bibinfo
  {author} {\bibfnamefont {J.}~\bibnamefont {Kirkman-Brown}},\ }\href@noop {}
  {\bibfield  {journal} {\bibinfo  {journal} {J Fluid Mech}\ }\textbf {\bibinfo
  {volume} {621}},\ \bibinfo {pages} {289} (\bibinfo {year}
  {2009}{\natexlab{a}})}\BibitemShut {NoStop}%
\bibitem [{\citenamefont {Gagnon}\ \emph {et~al.}(2013)\citenamefont {Gagnon},
  \citenamefont {Shen},\ and\ \citenamefont {Arratia}}]{Gagnon13}%
  \BibitemOpen
  \bibfield  {author} {\bibinfo {author} {\bibfnamefont {D.}~\bibnamefont
  {Gagnon}}, \bibinfo {author} {\bibfnamefont {X.}~\bibnamefont {Shen}}, \ and\
  \bibinfo {author} {\bibfnamefont {P.}~\bibnamefont {Arratia}},\ }\href@noop
  {} {\bibfield  {journal} {\bibinfo  {journal} {Europhys Lett}\ }\textbf
  {\bibinfo {volume} {104}},\ \bibinfo {pages} {14004} (\bibinfo {year}
  {2013})}\BibitemShut {NoStop}%
\bibitem [{\citenamefont {Fauci}\ and\ \citenamefont {Dillon}(2006)}]{Fauci06}%
  \BibitemOpen
  \bibfield  {author} {\bibinfo {author} {\bibfnamefont {L.}~\bibnamefont
  {Fauci}}\ and\ \bibinfo {author} {\bibfnamefont {R.}~\bibnamefont {Dillon}},\
  }\href@noop {} {\bibfield  {journal} {\bibinfo  {journal} {Annu Rev Fluid
  Mech}\ }\textbf {\bibinfo {volume} {38}},\ \bibinfo {pages} {371} (\bibinfo
  {year} {2006})}\BibitemShut {NoStop}%
\bibitem [{\citenamefont {Rutllant}\ \emph {et~al.}(2005)\citenamefont
  {Rutllant}, \citenamefont {Lopez-Bejar},\ and\ \citenamefont
  {Lopez-Gatius}}]{Rutllant05}%
  \BibitemOpen
  \bibfield  {author} {\bibinfo {author} {\bibfnamefont {J.}~\bibnamefont
  {Rutllant}}, \bibinfo {author} {\bibfnamefont {M.}~\bibnamefont
  {Lopez-Bejar}}, \ and\ \bibinfo {author} {\bibfnamefont {F.}~\bibnamefont
  {Lopez-Gatius}},\ }\href@noop {} {\bibfield  {journal} {\bibinfo  {journal}
  {Reprod Dom Anim}\ }\textbf {\bibinfo {volume} {40}},\ \bibinfo {pages} {79}
  (\bibinfo {year} {2005})}\BibitemShut {NoStop}%
\bibitem [{\citenamefont {Suarez}\ and\ \citenamefont
  {Pacey}(2006)}]{Suarez06}%
  \BibitemOpen
  \bibfield  {author} {\bibinfo {author} {\bibfnamefont {S.}~\bibnamefont
  {Suarez}}\ and\ \bibinfo {author} {\bibfnamefont {A.}~\bibnamefont {Pacey}},\
  }\href@noop {} {\bibfield  {journal} {\bibinfo  {journal} {Humn Reprod
  Update}\ }\textbf {\bibinfo {volume} {12}},\ \bibinfo {pages} {23} (\bibinfo
  {year} {2006})}\BibitemShut {NoStop}%
\bibitem [{\citenamefont {Celli}\ \emph {et~al.}(2009)\citenamefont {Celli},
  \citenamefont {Turner}, \citenamefont {Afdhal}, \citenamefont {Keates},
  \citenamefont {Ghiran}, \citenamefont {Kelly}, \citenamefont {Ewoldt},
  \citenamefont {McKinley}, \citenamefont {So}, \citenamefont {Erramilli},\
  and\ \citenamefont {Bansil}}]{Celli09}%
  \BibitemOpen
  \bibfield  {author} {\bibinfo {author} {\bibfnamefont {J.}~\bibnamefont
  {Celli}}, \bibinfo {author} {\bibfnamefont {B.}~\bibnamefont {Turner}},
  \bibinfo {author} {\bibfnamefont {N.}~\bibnamefont {Afdhal}}, \bibinfo
  {author} {\bibfnamefont {S.}~\bibnamefont {Keates}}, \bibinfo {author}
  {\bibfnamefont {I.}~\bibnamefont {Ghiran}}, \bibinfo {author} {\bibfnamefont
  {C.}~\bibnamefont {Kelly}}, \bibinfo {author} {\bibfnamefont
  {R.}~\bibnamefont {Ewoldt}}, \bibinfo {author} {\bibfnamefont
  {G.}~\bibnamefont {McKinley}}, \bibinfo {author} {\bibfnamefont
  {P.}~\bibnamefont {So}}, \bibinfo {author} {\bibfnamefont {S.}~\bibnamefont
  {Erramilli}}, \ and\ \bibinfo {author} {\bibfnamefont {R.}~\bibnamefont
  {Bansil}},\ }\href@noop {} {\bibfield  {journal} {\bibinfo  {journal} {Proc
  Natl Acad Sci USA}\ }\textbf {\bibinfo {volume} {106}},\ \bibinfo {pages}
  {1431} (\bibinfo {year} {2009})}\BibitemShut {NoStop}%
\bibitem [{\citenamefont {Flemming}\ and\ \citenamefont
  {Wingender}(2010)}]{Flemming10}%
  \BibitemOpen
  \bibfield  {author} {\bibinfo {author} {\bibfnamefont {H.}~\bibnamefont
  {Flemming}}\ and\ \bibinfo {author} {\bibfnamefont {J.}~\bibnamefont
  {Wingender}},\ }\href@noop {} {\bibfield  {journal} {\bibinfo  {journal}
  {Nature Rev Microbiol}\ }\textbf {\bibinfo {volume} {8}},\ \bibinfo {pages}
  {623} (\bibinfo {year} {2010})}\BibitemShut {NoStop}%
\bibitem [{\citenamefont {Berg}\ and\ \citenamefont {Turner}(1979)}]{Berg79}%
  \BibitemOpen
  \bibfield  {author} {\bibinfo {author} {\bibfnamefont {H.}~\bibnamefont
  {Berg}}\ and\ \bibinfo {author} {\bibfnamefont {L.}~\bibnamefont {Turner}},\
  }\href@noop {} {\bibfield  {journal} {\bibinfo  {journal} {Nature}\ }\textbf
  {\bibinfo {volume} {278}},\ \bibinfo {pages} {349} (\bibinfo {year}
  {1979})}\BibitemShut {NoStop}%
\bibitem [{\citenamefont {Schneider}\ and\ \citenamefont
  {Doetsch}(1974)}]{Schneider74}%
  \BibitemOpen
  \bibfield  {author} {\bibinfo {author} {\bibfnamefont {W.}~\bibnamefont
  {Schneider}}\ and\ \bibinfo {author} {\bibfnamefont {R.}~\bibnamefont
  {Doetsch}},\ }\href@noop {} {\bibfield  {journal} {\bibinfo  {journal} {J
  Bacteriol}\ }\textbf {\bibinfo {volume} {117}},\ \bibinfo {pages} {696}
  (\bibinfo {year} {1974})}\BibitemShut {NoStop}%
\bibitem [{\citenamefont {Smith}\ \emph
  {et~al.}(2009{\natexlab{b}})\citenamefont {Smith}, \citenamefont {Gaffney},
  \citenamefont {Gadelha}, \citenamefont {Kapur},\ and\ \citenamefont
  {Kirkman-Brown}}]{Smith09b}%
  \BibitemOpen
  \bibfield  {author} {\bibinfo {author} {\bibfnamefont {D.}~\bibnamefont
  {Smith}}, \bibinfo {author} {\bibfnamefont {E.}~\bibnamefont {Gaffney}},
  \bibinfo {author} {\bibfnamefont {H.}~\bibnamefont {Gadelha}}, \bibinfo
  {author} {\bibfnamefont {N.}~\bibnamefont {Kapur}}, \ and\ \bibinfo {author}
  {\bibfnamefont {J.}~\bibnamefont {Kirkman-Brown}},\ }\href@noop {} {\bibfield
   {journal} {\bibinfo  {journal} {Cell Motil Cytoskel}\ }\textbf {\bibinfo
  {volume} {66}},\ \bibinfo {pages} {220} (\bibinfo {year}
  {2009}{\natexlab{b}})}\BibitemShut {NoStop}%
\bibitem [{\citenamefont {Suarez}\ and\ \citenamefont {Dai}(1992)}]{Suarez92}%
  \BibitemOpen
  \bibfield  {author} {\bibinfo {author} {\bibfnamefont {S.}~\bibnamefont
  {Suarez}}\ and\ \bibinfo {author} {\bibfnamefont {X.}~\bibnamefont {Dai}},\
  }\href@noop {} {\bibfield  {journal} {\bibinfo  {journal} {Biol Reprod}\
  }\textbf {\bibinfo {volume} {46}},\ \bibinfo {pages} {686} (\bibinfo {year}
  {1992})}\BibitemShut {NoStop}%
\bibitem [{\citenamefont {Taylor}(1951)}]{Taylor51}%
  \BibitemOpen
  \bibfield  {author} {\bibinfo {author} {\bibfnamefont {G.}~\bibnamefont
  {Taylor}},\ }\href@noop {} {\bibfield  {journal} {\bibinfo  {journal} {Proc
  Roy Soc Lond Ser A}\ }\textbf {\bibinfo {volume} {209}},\ \bibinfo {pages}
  {447} (\bibinfo {year} {1951})}\BibitemShut {NoStop}%
\bibitem [{\citenamefont {Taylor}(1952)}]{Taylor52}%
  \BibitemOpen
  \bibfield  {author} {\bibinfo {author} {\bibfnamefont {G.}~\bibnamefont
  {Taylor}},\ }\href@noop {} {\bibfield  {journal} {\bibinfo  {journal} {Proc
  Roy Soc Lond Ser A}\ }\textbf {\bibinfo {volume} {211}},\ \bibinfo {pages}
  {225} (\bibinfo {year} {1952})}\BibitemShut {NoStop}%
\bibitem [{\citenamefont {Kosa}\ \emph {et~al.}(2010)\citenamefont {Kosa},
  \citenamefont {Shoham},\ and\ \citenamefont {Haber}}]{Kosa10}%
  \BibitemOpen
  \bibfield  {author} {\bibinfo {author} {\bibfnamefont {G.}~\bibnamefont
  {Kosa}}, \bibinfo {author} {\bibfnamefont {M.}~\bibnamefont {Shoham}}, \ and\
  \bibinfo {author} {\bibfnamefont {S.}~\bibnamefont {Haber}},\ }\href@noop {}
  {\bibfield  {journal} {\bibinfo  {journal} {Phys Fluids}\ }\textbf {\bibinfo
  {volume} {22}},\ \bibinfo {pages} {083101} (\bibinfo {year}
  {2010})}\BibitemShut {NoStop}%
\bibitem [{\citenamefont {Sauzade}\ \emph {et~al.}(2012)\citenamefont
  {Sauzade}, \citenamefont {Elfring},\ and\ \citenamefont {Lauga}}]{Sauzade12}%
  \BibitemOpen
  \bibfield  {author} {\bibinfo {author} {\bibfnamefont {M.}~\bibnamefont
  {Sauzade}}, \bibinfo {author} {\bibfnamefont {G.}~\bibnamefont {Elfring}}, \
  and\ \bibinfo {author} {\bibfnamefont {E.}~\bibnamefont {Lauga}},\
  }\href@noop {} {\bibfield  {journal} {\bibinfo  {journal} {Physica D}\
  }\textbf {\bibinfo {volume} {240}},\ \bibinfo {pages} {1567} (\bibinfo {year}
  {2012})}\BibitemShut {NoStop}%
\bibitem [{\citenamefont {Fu}\ \emph {et~al.}(2010)\citenamefont {Fu},
  \citenamefont {Shenoy},\ and\ \citenamefont {Powers}}]{Fu10}%
  \BibitemOpen
  \bibfield  {author} {\bibinfo {author} {\bibfnamefont {H.}~\bibfnamefont {C.}~\bibnamefont
  {Fu}}, \bibinfo {author} {\bibfnamefont {V.}~\bibnamefont {Shenoy}}, \ and\
  \bibinfo {author} {\bibfnamefont {T.}~\bibfnamefont {R.}~\bibnamefont {Powers}},\ }\href@noop {}
  {\bibfield  {journal} {\bibinfo  {journal} {Europhys Lett}\ }\textbf
  {\bibinfo {volume} {91}} (\bibinfo {year} {2010})}\BibitemShut {NoStop}%
\bibitem [{\citenamefont {Du}\ \emph {et~al.}(2012)\citenamefont {Du},
  \citenamefont {Keener}, \citenamefont {Guy},\ and\ \citenamefont
  {Fogelson}}]{Du12}%
  \BibitemOpen
  \bibfield  {author} {\bibinfo {author} {\bibfnamefont {J.}~\bibnamefont
  {Du}}, \bibinfo {author} {\bibfnamefont {J.}~\bibfnamefont {P.}~\bibnamefont {Keener}}, \bibinfo
  {author} {\bibfnamefont {R.}~\bibfnamefont {D.}~\bibnamefont {Guy}}, \ and\ \bibinfo {author}
  {\bibfnamefont {A.}~\bibfnamefont {L.}~\bibnamefont {Fogelson}},\ }\href@noop {} {\bibfield
  {journal} {\bibinfo  {journal} {Phys Rev E}\ }\textbf {\bibinfo {volume}
  {85}},\ \bibinfo {pages} {036304} (\bibinfo {year} {2012})}\BibitemShut
  {NoStop}%
\bibitem [{\citenamefont {Magariyama}\ and\ \citenamefont
  {S}(2002)}]{Magariyama02}%
  \BibitemOpen
  \bibfield  {author} {\bibinfo {author} {\bibfnamefont {Y.}~\bibnamefont
  {Magariyama}}\ and\ \bibinfo {author} {\bibfnamefont {K.}~\bibnamefont {S}},\
  }\href@noop {} {\bibfield  {journal} {\bibinfo  {journal} {Biophys J}\
  }\textbf {\bibinfo {volume} {83}},\ \bibinfo {pages} {733} (\bibinfo {year}
  {2002})}\BibitemShut {NoStop}%
\bibitem [{\citenamefont {Montenegro-Johnson}\ \emph
  {et~al.}(2013)\citenamefont {Montenegro-Johnson}, \citenamefont {Smith},\
  and\ \citenamefont {Loghin}}]{Montenegro13}%
  \BibitemOpen
  \bibfield  {author} {\bibinfo {author} {\bibfnamefont {T.}~\bibnamefont
  {Montenegro-Johnson}}, \bibinfo {author} {\bibfnamefont {D.}~\bibnamefont
  {Smith}}, \ and\ \bibinfo {author} {\bibfnamefont {D.}~\bibnamefont
  {Loghin}},\ }\href@noop {} {\bibfield  {journal} {\bibinfo  {journal} {Phys
  Fluids}\ }\textbf {\bibinfo {volume} {25}},\ \bibinfo {pages} {081903}
  (\bibinfo {year} {2013})}\BibitemShut {NoStop}%
\bibitem [{\citenamefont {Dasgupta}\ \emph {et~al.}(2013)\citenamefont
  {Dasgupta}, \citenamefont {Liu}, \citenamefont {Fu}, \citenamefont {Berhanu},
  \citenamefont {Breuer}, \citenamefont {Powers},\ and\ \citenamefont
  {Kudrolli}}]{Dasgupta13}%
  \BibitemOpen
  \bibfield  {author} {\bibinfo {author} {\bibfnamefont {M.}~\bibnamefont
  {Dasgupta}}, \bibinfo {author} {\bibfnamefont {B.}~\bibnamefont {Liu}},
  \bibinfo {author} {\bibfnamefont {H.}~\bibfnamefont {C.}~\bibnamefont {Fu}}, \bibinfo {author}
  {\bibfnamefont {M.}~\bibnamefont {Berhanu}}, \bibinfo {author} {\bibfnamefont
  {K.}~\bibfnamefont {S.}~\bibnamefont {Breuer}}, \bibinfo {author} {\bibfnamefont
  {T.}~\bibfnamefont {R.}~\bibnamefont {Powers}}, \ and\ \bibinfo {author} {\bibfnamefont
  {A.}~\bibnamefont {Kudrolli}},\ }\href@noop {} {\bibfield  {journal}
  {\bibinfo  {journal} {Phys Rev E}\ }\textbf {\bibinfo {volume} {87}},\
  \bibinfo {pages} {013015} (\bibinfo {year} {2013})}\BibitemShut {NoStop}%
\bibitem [{\citenamefont {Fu}\ \emph {et~al.}(2007)\citenamefont {Fu},
  \citenamefont {Powers},\ and\ \citenamefont {Wolgemuth}}]{Fu07}%
  \BibitemOpen
  \bibfield  {author} {\bibinfo {author} {\bibfnamefont {H.}~\bibfnamefont {C.}~\bibnamefont
  {Fu}}, \bibinfo {author} {\bibfnamefont {T.}~\bibfnamefont {R.}~\bibnamefont {Powers}}, \ and\
  \bibinfo {author} {\bibfnamefont {C.}~\bibfnamefont {W.}~\bibnamefont {Wolgemuth}},\ }\href@noop
  {} {\bibfield  {journal} {\bibinfo  {journal} {Phys Rev Lett}\ }\textbf
  {\bibinfo {volume} {99}},\ \bibinfo {pages} {258101} (\bibinfo {year}
  {2007})}\BibitemShut {NoStop}%
\bibitem [{\citenamefont {Fu}\ \emph {et~al.}(2009)\citenamefont {Fu},
  \citenamefont {Wolgemuth},\ and\ \citenamefont {Powers}}]{Fu09}%
  \BibitemOpen
  \bibfield  {author} {\bibinfo {author} {\bibfnamefont {H.}~\bibfnamefont {C.}~\bibnamefont
  {Fu}}, \bibinfo {author} {\bibfnamefont {C.}~\bibfnamefont {W.}~\bibnamefont {Wolgemuth}}, \
  and\ \bibinfo {author} {\bibfnamefont {T.}~\bibfnamefont {R.}~\bibnamefont {Powers}},\
  }\href@noop {} {\bibfield  {journal} {\bibinfo  {journal} {Phys Fluids}\
  }\textbf {\bibinfo {volume} {21}},\ \bibinfo {pages} {033102} (\bibinfo
  {year} {2009})}\BibitemShut {NoStop}%
\bibitem [{\citenamefont {Lauga}(2007)}]{Lauga07}%
  \BibitemOpen
  \bibfield  {author} {\bibinfo {author} {\bibfnamefont {E.}~\bibnamefont
  {Lauga}},\ }\href@noop {} {\bibfield  {journal} {\bibinfo  {journal} {Phys
  Fluids}\ }\textbf {\bibinfo {volume} {19}},\ \bibinfo {pages} {083104}
  (\bibinfo {year} {2007})}\BibitemShut {NoStop}%
\bibitem [{\citenamefont {Teran}\ \emph {et~al.}(2010)\citenamefont {Teran},
  \citenamefont {Fauci},\ and\ \citenamefont {Shelley}}]{Teran10}%
  \BibitemOpen
  \bibfield  {author} {\bibinfo {author} {\bibfnamefont {J.}~\bibnamefont
  {Teran}}, \bibinfo {author} {\bibfnamefont {L.}~\bibnamefont {Fauci}}, \ and\
  \bibinfo {author} {\bibfnamefont {M.}~\bibnamefont {Shelley}},\ }\href@noop
  {} {\bibfield  {journal} {\bibinfo  {journal} {Phys Rev Lett}\ }\textbf
  {\bibinfo {volume} {104}},\ \bibinfo {pages} {038101} (\bibinfo {year}
  {2010})}\BibitemShut {NoStop}%
\bibitem [{\citenamefont {Thomases}\ and\ \citenamefont
  {Guy}(2014)}]{Thomases14}%
  \BibitemOpen
  \bibfield  {author} {\bibinfo {author} {\bibfnamefont {B.}~\bibnamefont
  {Thomases}}\ and\ \bibinfo {author} {\bibfnamefont {R.}~\bibfnamefont {D.}~\bibnamefont {Guy}},\
  }\href@noop {} {\bibfield  {journal} {\bibinfo  {journal} {Phys Rev Lett}\
  }\textbf {\bibinfo {volume} {113}},\ \bibinfo {pages} {098102} (\bibinfo
  {year} {2014})}\BibitemShut {NoStop}%
\bibitem [{\citenamefont {Brinkman}(1947)}]{Brinkman47}%
  \BibitemOpen
  \bibfield  {author} {\bibinfo {author} {\bibfnamefont {H.}~\bibnamefont
  {Brinkman}},\ }\href@noop {} {\bibfield  {journal} {\bibinfo  {journal} {Appl
  Sci Res}\ }\textbf {\bibinfo {volume} {A1}},\ \bibinfo {pages} {27} (\bibinfo
  {year} {1947})}\BibitemShut {NoStop}%
\bibitem [{\citenamefont {Koplik}\ \emph {et~al.}(1983)\citenamefont {Koplik},
  \citenamefont {Levine},\ and\ \citenamefont {Zee}}]{Koplik83}%
  \BibitemOpen
  \bibfield  {author} {\bibinfo {author} {\bibfnamefont {J.}~\bibnamefont
  {Koplik}}, \bibinfo {author} {\bibfnamefont {H.}~\bibnamefont {Levine}}, \
  and\ \bibinfo {author} {\bibfnamefont {A.}~\bibnamefont {Zee}},\ }\href@noop
  {} {\bibfield  {journal} {\bibinfo  {journal} {Phys Fluids}\ }\textbf
  {\bibinfo {volume} {26}},\ \bibinfo {pages} {2864} (\bibinfo {year}
  {1983})}\BibitemShut {NoStop}%
\bibitem [{\citenamefont {Auriault}(2009)}]{Auriault09}%
  \BibitemOpen
  \bibfield  {author} {\bibinfo {author} {\bibfnamefont {J.}~\bibnamefont
  {Auriault}},\ }\href@noop {} {\bibfield  {journal} {\bibinfo  {journal}
  {Transp Porous Med}\ }\textbf {\bibinfo {volume} {79}},\ \bibinfo {pages}
  {215} (\bibinfo {year} {2009})}\BibitemShut {NoStop}%
\bibitem [{\citenamefont {Howells}(1974)}]{Howells74}%
  \BibitemOpen
  \bibfield  {author} {\bibinfo {author} {\bibfnamefont {I.}~\bibnamefont
  {Howells}},\ }\href@noop {} {\bibfield  {journal} {\bibinfo  {journal} {J
  Fluid Mech}\ }\textbf {\bibinfo {volume} {64}},\ \bibinfo {pages} {449}
  (\bibinfo {year} {1974})}\BibitemShut {NoStop}%
\bibitem [{\citenamefont {Spielman}\ and\ \citenamefont
  {Goren}(1968)}]{Spielman}%
  \BibitemOpen
  \bibfield  {author} {\bibinfo {author} {\bibfnamefont {L.}~\bibnamefont
  {Spielman}}\ and\ \bibinfo {author} {\bibfnamefont {S.}~\bibnamefont
  {Goren}},\ }\href@noop {} {\bibfield  {journal} {\bibinfo  {journal} {Env
  Science Tech}\ }\textbf {\bibinfo {volume} {1}},\ \bibinfo {pages} {279}
  (\bibinfo {year} {1968})}\BibitemShut {NoStop}%
\bibitem [{\citenamefont {Durlofsky}\ and\ \citenamefont
  {Brady}(1987)}]{Durlofsky87}%
  \BibitemOpen
  \bibfield  {author} {\bibinfo {author} {\bibfnamefont {L.}~\bibnamefont
  {Durlofsky}}\ and\ \bibinfo {author} {\bibfnamefont {J.}~\bibnamefont
  {Brady}},\ }\href@noop {} {\bibfield  {journal} {\bibinfo  {journal} {Physics
  of Fluids}\ }\textbf {\bibinfo {volume} {30}},\ \bibinfo {pages} {3329}
  (\bibinfo {year} {1987})}\BibitemShut {NoStop}%
\bibitem [{\citenamefont {Leshansky}(2009)}]{Leshansky09}%
  \BibitemOpen
  \bibfield  {author} {\bibinfo {author} {\bibfnamefont {A.}~\bibfnamefont {M.}~\bibnamefont
  {Leshansky}},\ }\href@noop {} {\bibfield  {journal} {\bibinfo  {journal}
  {Phys Rev E}\ }\textbf {\bibinfo {volume} {80}},\ \bibinfo {pages} {051911}
  (\bibinfo {year} {2009})}\BibitemShut {NoStop}%
\bibitem [{\citenamefont {Happel}\ and\ \citenamefont
  {Brenner}(1983)}]{Happel83}%
  \BibitemOpen
  \bibfield  {author} {\bibinfo {author} {\bibfnamefont {J.}~\bibnamefont
  {Happel}}\ and\ \bibinfo {author} {\bibfnamefont {H.}~\bibnamefont
  {Brenner}},\ }\href@noop {} {\emph {\bibinfo {title} {Low {R}eynolds number
  hydrodynamics}}},\ \bibinfo {edition} {2nd}\ ed.\ (\bibinfo  {publisher}
  {Martinus Nijhoff Publishers},\ \bibinfo {year} {1983})\ \bibinfo {note} {ch
  3, Section 3}\BibitemShut {NoStop}%
\bibitem [{\citenamefont {Olver}(1972)}]{Milton72}%
  \BibitemOpen
  \bibfield  {author} {\bibinfo {author} {\bibfnamefont {F.}~\bibnamefont
  {Olver}},\ }in\ \href@noop {} {\emph {\bibinfo {booktitle} {Handbook of
  mathematical functions: with formals, graphs, and mathematical tables}}},\
  \bibinfo {editor} {edited by\ \bibinfo {editor} {\bibfnamefont
  {M.}~\bibnamefont {Abramowitz}}\ and\ \bibinfo {editor} {\bibfnamefont
  {A.}~\bibnamefont {Stegun}}}\ (\bibinfo  {publisher} {Dover Publications},\
  \bibinfo {address} {Minneola, N.Y.},\ \bibinfo {year} {1972})\BibitemShut
  {NoStop}%
\bibitem [{\citenamefont {Drummond}(1966)}]{Drummond66}%
  \BibitemOpen
  \bibfield  {author} {\bibinfo {author} {\bibfnamefont {J.}~\bibnamefont
  {Drummond}},\ }\href@noop {} {\bibfield  {journal} {\bibinfo  {journal} {J
  Fluid Mech}\ }\textbf {\bibinfo {volume} {25}},\ \bibinfo {pages} {787}
  (\bibinfo {year} {1966})}\BibitemShut {NoStop}%
\bibitem [{\citenamefont {Saltzman}\ \emph {et~al.}(1994)\citenamefont
  {Saltzman}, \citenamefont {Radomsky}, \citenamefont {Whaley},\ and\
  \citenamefont {Cone}}]{saltzman1994antibody}%
  \BibitemOpen
  \bibfield  {author} {\bibinfo {author} {\bibfnamefont {W.~M.}\ \bibnamefont
  {Saltzman}}, \bibinfo {author} {\bibfnamefont {M.~L.}\ \bibnamefont
  {Radomsky}}, \bibinfo {author} {\bibfnamefont {K.~J.}\ \bibnamefont
  {Whaley}}, \ and\ \bibinfo {author} {\bibfnamefont {R.~A.}\ \bibnamefont
  {Cone}},\ }\href@noop {} {\bibfield  {journal} {\bibinfo  {journal} {Biophys
  J}\ }\textbf {\bibinfo {volume} {66}},\ \bibinfo {pages} {508} (\bibinfo
  {year} {1994})}\BibitemShut {NoStop}%
\bibitem [{\citenamefont {Hafez}\ and\ \citenamefont
  {Kenemans}(2012)}]{hafez2012atlas}%
  \BibitemOpen
  \bibfield  {author} {\bibinfo {author} {\bibfnamefont {E.}~\bibnamefont
  {Hafez}}\ and\ \bibinfo {author} {\bibfnamefont {P.}~\bibnamefont
  {Kenemans}},\ }\href@noop {} {\emph {\bibinfo {title} {Atlas of Human
  Reproduction: By Scanning Electron Microscopy}}}\ (\bibinfo  {publisher}
  {Springer Science \& Business Media},\ \bibinfo {year} {2012})\BibitemShut
  {NoStop}%
\bibitem [{\citenamefont {Bahr}\ and\ \citenamefont
  {Zeitler}(1964)}]{bahr1964study}%
  \BibitemOpen
  \bibfield  {author} {\bibinfo {author} {\bibfnamefont {G.~F.}\ \bibnamefont
  {Bahr}}\ and\ \bibinfo {author} {\bibfnamefont {E.}~\bibnamefont {Zeitler}},\
  }\href@noop {} {\bibfield  {journal} {\bibinfo  {journal} {J Cell Biol}\
  }\textbf {\bibinfo {volume} {21}},\ \bibinfo {pages} {175} (\bibinfo {year}
  {1964})}\BibitemShut {NoStop}%
\bibitem [{\citenamefont {Bloodgood}(2013)}]{bloodgood2013ciliary}%
  \BibitemOpen
  \bibfield  {author} {\bibinfo {author} {\bibfnamefont {R.}~\bibnamefont
  {Bloodgood}},\ }\href@noop {} {\emph {\bibinfo {title} {Ciliary and flagellar
  membranes}}}\ (\bibinfo  {publisher} {Springer Science \& Business Media},\
  \bibinfo {year} {2013})\BibitemShut {NoStop}%
\bibitem [{\citenamefont {Aguilar}\ and\ \citenamefont
  {Reyley}(2005)}]{Aguilar05}%
  \BibitemOpen
  \bibfield  {author} {\bibinfo {author} {\bibfnamefont {J.}~\bibnamefont
  {Aguilar}}\ and\ \bibinfo {author} {\bibfnamefont {M.}~\bibnamefont
  {Reyley}},\ }\href@noop {} {\bibfield  {journal} {\bibinfo  {journal} {Anim
  Reprod}\ }\textbf {\bibinfo {volume} {2}},\ \bibinfo {pages} {91} (\bibinfo
  {year} {2005})}\BibitemShut {NoStop}%
\bibitem [{\citenamefont {Cortez}\ \emph {et~al.}(2010)\citenamefont {Cortez},
  \citenamefont {Cummins}, \citenamefont {Leiderman},\ and\ \citenamefont
  {Varela}}]{Cortez10}%
  \BibitemOpen
  \bibfield  {author} {\bibinfo {author} {\bibfnamefont {R.}~\bibnamefont
  {Cortez}}, \bibinfo {author} {\bibfnamefont {B.}~\bibnamefont {Cummins}},
  \bibinfo {author} {\bibfnamefont {K.}~\bibnamefont {Leiderman}}, \ and\
  \bibinfo {author} {\bibfnamefont {D.}~\bibnamefont {Varela}},\ }\href@noop {}
  {\bibfield  {journal} {\bibinfo  {journal} {J Comp Phys}\ }\textbf {\bibinfo
  {volume} {229}},\ \bibinfo {pages} {7609} (\bibinfo {year}
  {2010})}\BibitemShut {NoStop}%
\bibitem [{\citenamefont {Cortez}(2001)}]{Cortez01}%
  \BibitemOpen
  \bibfield  {author} {\bibinfo {author} {\bibfnamefont {R.}~\bibnamefont
  {Cortez}},\ }\href@noop {} {\bibfield  {journal} {\bibinfo  {journal} {SIAM J
  Sci Comput}\ }\textbf {\bibinfo {volume} {23}},\ \bibinfo {pages} {1204}
  (\bibinfo {year} {2001})}\BibitemShut {NoStop}%
\bibitem [{\citenamefont {Cortez}\ \emph {et~al.}(2005)\citenamefont {Cortez},
  \citenamefont {Fauci},\ and\ \citenamefont {Medovikov}}]{Cortez05}%
  \BibitemOpen
  \bibfield  {author} {\bibinfo {author} {\bibfnamefont {R.}~\bibnamefont
  {Cortez}}, \bibinfo {author} {\bibfnamefont {L.}~\bibnamefont {Fauci}}, \
  and\ \bibinfo {author} {\bibfnamefont {A.}~\bibnamefont {Medovikov}},\
  }\href@noop {} {\bibfield  {journal} {\bibinfo  {journal} {Phys Fluids}\
  }\textbf {\bibinfo {volume} {17}},\ \bibinfo {pages} {0315041} (\bibinfo
  {year} {2005})}\BibitemShut {NoStop}%
\bibitem [{\citenamefont {Pozrikidis}(1989)}]{Pozrikidis1989a}%
  \BibitemOpen
  \bibfield  {author} {\bibinfo {author} {\bibfnamefont {C.}~\bibnamefont
  {Pozrikidis}},\ }\href@noop {} {\bibfield  {journal} {\bibinfo  {journal}
  {Phys Fluids A}\ }\textbf {\bibinfo {volume} {1}},\ \bibinfo {pages} {1508}
  (\bibinfo {year} {1989})}\BibitemShut {NoStop}%
\bibitem [{\citenamefont {Jung}\ \emph {et~al.}(2007)\citenamefont {Jung},
  \citenamefont {Mareck}, \citenamefont {Fauci},\ and\ \citenamefont
  {Shelly}}]{JungFauci07}%
  \BibitemOpen
  \bibfield  {author} {\bibinfo {author} {\bibfnamefont {S.}~\bibnamefont
  {Jung}}, \bibinfo {author} {\bibfnamefont {K.}~\bibnamefont {Mareck}},
  \bibinfo {author} {\bibfnamefont {L.}~\bibnamefont {Fauci}}, \ and\ \bibinfo
  {author} {\bibfnamefont {M.~J.}\ \bibnamefont {Shelly}},\ }\href@noop {}
  {\bibfield  {journal} {\bibinfo  {journal} {Phys Fluids}\ }\textbf {\bibinfo
  {volume} {19}},\ \bibinfo {pages} {103105} (\bibinfo {year}
  {2007})}\BibitemShut {NoStop}%
\bibitem [{\citenamefont {Sanders}(2009)}]{Sanders09}%
  \BibitemOpen
  \bibfield  {author} {\bibinfo {author} {\bibfnamefont {L.}~\bibnamefont
  {Sanders}},\ }\href@noop {} {\bibfield  {journal} {\bibinfo  {journal}
  {Science News}\ }\textbf {\bibinfo {volume} {176}},\ \bibinfo {pages} {22}
  (\bibinfo {year} {2009})}\BibitemShut {NoStop}%
\bibitem [{\citenamefont {Olson}\ and\ \citenamefont
  {Leiderman}(2015)}]{Olson15a}%
  \BibitemOpen
  \bibfield  {author} {\bibinfo {author} {\bibfnamefont {S.}~\bibfnamefont {D.}~\bibnamefont
  {Olson}}\ and\ \bibinfo {author} {\bibfnamefont {K.}~\bibnamefont
  {Leiderman}},\ }\href@noop {} {\bibfield  {journal} {\bibinfo  {journal} {J
  Aero Aqua Bio-mech}\ }\textbf {\bibinfo {volume} {4}},\ \bibinfo {pages} {12}
  (\bibinfo {year} {2015})}\BibitemShut {NoStop}%
\bibitem [{\citenamefont {Babcock}\ \emph {et~al.}(2014)\citenamefont
  {Babcock}, \citenamefont {Wandernoth},\ and\ \citenamefont
  {Wennemuth}}]{Babcock14}%
  \BibitemOpen
  \bibfield  {author} {\bibinfo {author} {\bibfnamefont {D.}~\bibnamefont
  {Babcock}}, \bibinfo {author} {\bibfnamefont {P.}~\bibnamefont {Wandernoth}},
  \ and\ \bibinfo {author} {\bibfnamefont {G.}~\bibnamefont {Wennemuth}},\
  }\href@noop {} {\bibfield  {journal} {\bibinfo  {journal} {BMC Biol}\
  }\textbf {\bibinfo {volume} {12}} (\bibinfo {year} {2014})}\BibitemShut
  {NoStop}%
\end{thebibliography}
%

\end{document}